\documentclass[nofootinbib,preprint,floatfix,superscriptaddress,showpacs,tightenlines]{revtex4}
\usepackage{graphicx}
\usepackage{amsfonts}
\usepackage{amssymb}
\usepackage{amsmath}
\usepackage{latexsym}

\begin{document}

\preprint{gr-qc/0303115}

\title{Brane classical and quantum cosmology from an effective action}

\author{Sanjeev S.~Seahra}
\email{ssseahra@uwaterloo.ca}
\affiliation{Department of Physics,
University of Waterloo, Waterloo, Ontario, N2L 3G1, Canada}

\author{H.~R.~Sepangi}
\email{hr-sepangi@cc.sbu.ac.ir}%
\affiliation{Department of Physics, University of Waterloo,
Waterloo, Ontario, N2L 3G1, Canada}%
\affiliation{Department of Physics, Shahid Beheshti University,
Evin, Tehran 19839 Iran.}

\author{J.~Ponce de Leon}
\email{jponce@upracd.upr.clu.edu}%
\affiliation{Centre for Gravitation and Fundamental Metrology,
VNIIMS, 3-1 M. Ulyanovoi Str., 117313 Moscow, Russia}%
\affiliation{Laboratory of Theoretical Physics, Department of
Physics, University of Puerto Rico, P.O. Box 23343, San Juan, PR
00931, USA.}

\date{July 18, 2003}

\setlength\arraycolsep{2pt}
\newcommand*{\di}{\partial}
\newcommand*{\V}{{\mathcal V}^{(k)}_d}
\newcommand*{\volume}{\sqrt{\sigma^{(k,d)}}}
\newcommand*{\OneTwo}{{(1,2)}}
\newcommand*{\onetwo}{{1,2}}
\newcommand*{\Lm}{{\mathcal L}_m}
\newcommand*{\Hm}{{\mathcal H}_m}
\newcommand*{\hatHm}{\hat{\mathcal H}_m}
\newcommand*{\Ldust}{{\mathcal L}_\mathrm{dust}}
\newcommand*{\maxsym}{{\mathbb S}_d^{(k)}}

\begin{abstract}

Motivated by the Randall-Sundrum brane-world scenario, we discuss
the classical and quantum dynamics of a $(d+1)$-dimensional
boundary wall between a pair of $(d+2)$-dimensional topological
Schwarzschild-AdS black holes.  We assume there are quite general
--- but not completely arbitrary --- matter fields living on the
boundary ``brane universe'' and its geometry is that of an
Friedmann-Lema\^{\i}tre-Robertson-Walker (FLRW) model.  The
effective action governing the model in the mini-superspace
approximation is derived.  We find that the presence of black hole
horizons in the bulk gives rise to a complex action for certain
classically allowed brane configurations, but that the imaginary
contribution plays no role in the equations of motion. Classical
and instanton brane trajectories are examined in general and for
special cases, and we find a subset of configuration space that is
not allowed at the classical or semi-classical level; these
correspond to spacelike branes carrying tachyonic matter. The
Hamiltonization and Dirac quantization of the model is then
performed for the general case; the latter involves the
manipulation of the Hamiltonian constraint before it is
transformed into an operator that annihilates physical state
vectors.  The ensuing covariant Wheeler-DeWitt equation is
examined at the semi-classical level, and we consider the possible
localization of the brane universe's wavefunction away from the
cosmological singularity.  This is easier to achieve for branes
with low density and/or spherical spatial sections.

\end{abstract}

\pacs{11.25.Wx, 04.70.Dy, 98.80.Qc}

\maketitle

%

\section{Introduction}\label{sec:intro}

The idea that our universe might be a 4-dimensional hypersurface
embedded in a higher-dimensional manifold is an old one with a
long history, as well as the subject of a considerable amount of
contemporary interest.  The primordial impetus for this line of
study comes from the work of Kaluza \cite{Kal21}, who showed that
one can obtain a classical unification of gravity and
electromagnetism by adding an extra dimension to spacetime (1921);
and Klein \cite{Kle26}, who suggested that extra dimensions have a
circular topology of very small radius and are hence unobservable
(1926).  The latter idea, the so-called ``compactification''
paradigm, came to dominate most approaches to higher-dimensional
physics, the most notable of which was early superstring theory.
However, a number of papers have appeared in the intervening years
that do not assume extra dimensions with compact topologies; early
examples include the works of Joseph \cite{Jos62}, Akama
\cite{Aka82}, Rubakov \& Shaposhnikov \cite{Rub83}, Visser
\cite{Vis85}, Gibbons \& Wiltshire \cite{Gib87}, and Antoniadis
\cite{Ant00}. A systematic and independent approach to the
5-dimensional, non-compact Kaluza-Klein scenario, known as
Space-Time-Matter (STM) theory, followed
\cite{Wes92,Wes96,Ove97,Wes99}. Then, in 1996 Horava \& Witten
showed that the compactification paradigm was not a prerequisite
of string theory with their discovery of an 11-dimensional theory
on the orbifold $\mathbb{R}^{10} \times S^1 / \mathbb{Z}_2$, which
is related to the 10-dimensional $E_8 \times E_8$ heterotic string
via dualities \cite{Hor96}. In this theory, standard model
interactions are confined to a lower-dimensional hypersurface,
known as a ``brane'', on which the endpoints of open strings
reside, while gravitation propagates in the higher-dimensional
bulk. This situation has come to be known as the ``braneworld
scenario.'' The works of Arkani-Hamed \emph{et al.}
\cite{Ark98a,Ant98,Ark98b} and Randall \& Sundrum (RS)
\cite{Ran99a,Ran99b}, which used non-compact extra dimensions to
address the hierarchy problem of particle physics and demonstrated
that the graviton ground state can be localized on a 3-brane in 5
dimensions, won a large following for the braneworld scenario and
a virtual flood of papers dealing with non-compact,
higher-dimensional models of the universe soon followed.

One particular type of braneworld model that has received much
attention concerns the idea that the universe could be a
4-dimensional boundary wall between a pair of 5-dimensional
Schwarzschild-anti deSitter (S-AdS${}_5$) or
Riesner-Nordstr\"om-anti deSitter black holes.\footnote{A critical
analysis of this and other types of cosmological brane world
models is given by Coule \cite{Cou01}.}  The Friedman equation
governing the classical dynamics of such a scenario has been
derived using Israel's junction conditions \cite{Isr66} for
arbitrary brane matter-content \cite{Kra99,Bar00}, and for the
case where the only matter energy on the brane is from its tension
or vacuum energy \cite{Sav01,Noj02,Gre02}.\footnote{We shall call
branes whose matter content consists solely of a cosmological
constant ``vacuum branes''.} There have been numerous studies of
the classical brane trajectories associated with such scenarios
that have found that the ``brane universe'' can exhibit
non-standard bouncing or cyclic behaviour
\cite{Cam01,Myu02,Muk02,Med02}.  Generalizations to
six-dimensional bulks have also been considered \cite{Cua03}.

The natural extension of this work is the problem of quantizing
these braneworld models.  Several authors have found it useful to
appeal to the well-known 4-dimensional formalism of quantum
cosmology to initiate studies of this issue, particularly
concerning the ``quantum birth'' of the universe
\cite{Vil82,Har83,Vil84,Lin84,Vil86,Rub84}.  The advantage of the
quantum cosmology approach are obvious: the mini-superspace
approximation allows one to pick a few of the system's degrees of
freedom to treat quantum mechanically while the others are
represented by their classical solutions.  However, there are a
number of well-catalogued problems associated with quantum
cosmology; including the problem of time \cite{Kuc91}, the
validity of the mini-superspace approximation, and the problem of
assigning appropriate boundary conditions to the wavefunction of
the universe \cite{Har97}.

Despite these difficulties, the canonical quantum cosmology for
the vacuum branes surrounding bulk black holes has been considered
from the point of view of an effective action \cite{Koy00,Bis03},
while the problem for a vacuum brane bounding pure AdS space has
also been  dealt with \cite{Anc00}.  The case where there is some
conformal field theory living on the brane has also been
considered for various bulk manifolds
\cite{Noj00aa,Noj00bb,Noj01aa}. Related to these studies are works
that consider the quantum creation (or decay) of brane universes
via saddle-point approximations to path integrals
\cite{Gor00,Ida01,Gre01} --- often in bulk manifolds other than
S-AdS${}_5$ --- as well as papers that consider the classical and
quantum dynamics of ``geodetic brane universes''
\cite{Kar02,Cor02,Cor03}.

It should be mentioned that quantum brane world models where the
bulk is sourced by black holes are related to 4-dimensional
problems other than canonical quantum cosmology.  Indeed, they
share many of the same features as the problem of the quantum
collapse of spherical matter shells, which has been rather
whimsically named ``quantum conchology'' by some authors
\cite{Haj92a,Haj92b,Fri97,Haj97,Kuc99,Cor01,Haj02}.  Also, the
problem associated with the quantum birth of a braneworld
sandwiched in between topological bulk black holes is almost the
inverse of the problem of the creation of 4-dimensional
topological black holes separated by a 3-dimensional domain wall
\cite{Man95,Cal96,Man98}.

The purpose of this paper is to analyze the classical and quantum
mechanics of a $d$-brane acting as a boundary between a pair of
``topological'' S-AdS${}_{(d+2)}$ black holes\footnote{The
adjective ``topological'' is meant to indicate that the geometry
of the $d$-surfaces of constant time and radius can be spherical,
flat, or hyperbolic.} from the point of view of an effective
action.  The treatment is designed to be as self-contained and
transparent as possible.  The action for our model in the
mini-superspace approximation is explicitly constructed from the
standard action of general relativity in Sec.~\ref{sec:action}.
One of the novel features of our analysis is that we allow for
arbitrary matter living on the brane, provided that the perfect
cosmological principle (PCP) is obeyed. The true dynamical
variables in our action are the brane radius and the matter field
configuration variables.  We also retain a gauge degree of freedom
in the form of the lapse function on the brane, which is
associated with transformations of the $(d+1)$-dimensional time
coordinate.  As a result, our effective action is
reparametrization invariant.  Another way in which our work
differs from previous efforts is that we pay special attention to
the behavior of the action as the brane crosses the bulk black
hole horizon --- if such a horizon exists --- and we will
demonstrate that even though the brane trajectory is perfectly
well behaved at the horizon, the action becomes complex valued.
This behaviour is reminiscent of the action governing the collapse
of thin matter shells in 4 dimensions.  We argue that a complex
action can be partly avoided by the addition of total time
derivatives to the Lagrangian in certain classically allowed parts
of configuration space --- which is the manifold spanned by the
system's coordinates and velocities
--- resulting in an piecewise-defined action.  However, we do
find a portion of configuration space where it is impossible to
make the action real.  When the brane is within this region, its
normal becomes timelike and comoving brane observers follow
spacelike paths. Hence we name this portion of configuration space
the ``tachyon region''.

Sec.~\ref{sec:classical} is devoted to the classical cosmology of
our model as derived from direct variation of the effective
action. We analyze the Friedman equation in the general situation
and determine the criterion for classically allowed and
classically forbidden regions.  We also derive the Newtonian limit
of the brane's equation of motion and show that it is nothing more
than an energy conservation equation for a thin-shell encircling a
central mass with zero total energy.  We then turn our attention
to a special case that is suitable for exact analysis.  In that
case, the bulk cosmological constant is set to zero and the
spatial sections of the brane are flat, while the matter on the
brane takes the form of the brane tension and a cosmological dust
fluid.\footnote{A streamlined version of Schutz's velocity
potential variational formalism for perfect fluids
\cite{Shu70,Shu71} --- like dust and vacuum energy --- is
developed in Appendix \ref{app:dust}.} We demonstrate that the
solutions of the Friedman equation exhibit exotic bounce and
crunch behaviour for negative mass bulk black holes; i.e., one
does not need a charged bulk black hole to avoid the cosmological
singularity. However, at least for this special case, one must
allow the energy conditions to be violated in the bulk. We also
look for classical brane trajectories that transverse the tachyon
region and find that only possibility is to allow the brane's
density to be imaginary.  This is not surprising; recall that only
point particles with imaginary mass can travel on spacelike
trajectories.

The Hamiltonization and quantization of the model is the subject
of Sec.~\ref{sec:quantum}.  In order to maintain a certain level
of rigor, we find that the reparametrization invariance of the
action and the general nature of the matter fields demand an
extended foray into Dirac's formalism dealing with the Hamiltonian
mechanics of constrained systems \cite{Dir64,Git90}.  The
piecewise nature of the action comes back to haunt us here; in
order to avoid the consideration of a complex phase space, we find
that it is necessary to define canonical momenta and constraints
in a piecewise fashion.  Continuity of the latter across the
horizon is resolved by rewriting the first-class Hamiltonian
constraint in an algebraically equivalent form and by making
minimal assumptions about the matter fields. We ultimately obtain
a continuous set of first-class constraints and Dirac bracket
structure suitable for Dirac quantization. The transformation of
the Hamiltonian constraint represents a significant departure from
previous studies \cite{Anc00,Bis03}, which is another ``twist''
that has a number of beneficial qualities.  The Wheeler-DeWitt
equation obtained upon quantization is shown to be equivalent to a
$(d+2)$-dimensional covariant wave equation --- which means that
it ought to be invariant under $(d+2)$-dimensional coordinate
transformations --- and reduces to a one-dimensional Schr\"odinger
equation that exhibits no pathological behaviour at the position
of the bulk horizon; this is in contrast to the wave equation
derived in \cite{Bis03}.  We then specialize to the case where the
brane matter consists of vacuum energy plus dust, and demonstrate
that for certain model parameters the wavefunction of the universe
can be localized away from the cosmological singularity by
potential barriers in the Wheeler-DeWitt equation.  The degree of
localization is characterized by the WKB tunnelling amplitude
through those same barriers, which is calculated explicitly for
certain model parameters.  We find that singularity avoidance is
more likely for branes with low matter density and spherical
spatial sections.

Finally Sec.~\ref{sec:summary} gives a summary of our results and
suggestions for future projects based on this work, of which there
are several.

\section{An effective action for the brane
world}\label{sec:action}

The model that we will be concerned with in this paper is as
follows:  Consider a $d$-brane $\Sigma$ that acts as a domain wall
between two bulk $N$-dimensional manifolds, where $N = d+2$. We
treat the embedding functions of the brane as the dynamical
degrees of freedom of the model, but we regard the components of
the two bulk metrics as fixed; i.e, we are considering a brane
propagating in a static background. The other dynamical degrees of
freedom in the model come from matter fields living on $\Sigma$,
about which we will make minimal assumptions.

The structure of the brane is taken to be that of an
$(d+1)$-dimensional FLRW model; i.e., $\Sigma = \mathbb{R} \times
\maxsym$, where $\maxsym$ is an $d$-dimensional Euclidean space of
constant curvature $k=-1,0,1$.\footnote{For example, when $k =1$
we have that $\maxsym$ is an $d$-sphere.}  We take $\theta = \{
\theta^1,\ldots,\theta^d\}$ to be a suitable coordinate system on
$\maxsym$ such that the metric is $\sigma^{(k,d)}_{ab}$, where
$a,b = 1 \ldots d$.  A necessary assumption for a well defined
action of our model is that $\maxsym$ is globally compact; i.e.,
if $k=0$ we take $\maxsym$ to be an $d$-torus and if $k=-1$ we
take $\maxsym$ to be a compact $d$-hyperboloid. The finite
$d$-dimensional volume of the unit radius submanifold $\maxsym$ is
then given by
\begin{equation}
    \V = \oint\limits_{\maxsym} d^d\theta \, \volume,
\end{equation}
where $\sigma^{(k,d)} = \det\sigma^{(k,d)}_{ab}$.  Our work will
not depend on the actual value of $\V$ --- other than the fact
that it is finite --- so we do not need to specify the periodicity
of the $\theta^a$ coordinates in the $k = 0$ or $-1$ cases (when
$k=1$ we take $\theta^a$ to be the standard angular coordinates on
an $d$-sphere).

\begin{figure}
\includegraphics{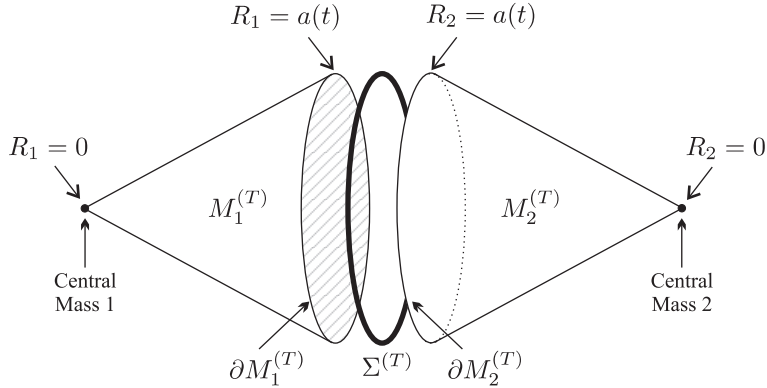}
\caption{A pictorial representation of a spatial slice of our
model.  The $(T)$ superscripts are meant to convey that we are
showing the $T=$ constant surfaces of the relevant manifolds.  We
have suppressed $d-1$ dimensions on the $\maxsym$ submanifolds so
that they appear as vertical circles and $M_\onetwo^{(T)}$ appear
as 2-surfaces.  The finite amount of space between $\di
M_1^{(T)}$, $\Sigma^{(T)}$, and $\di M_2^{(T)}$ is included to
ease with visualization; in actuality those three $d$-surfaces are
coincident. Note that the spatial slice is compact for finite
$a(t)$ and that each of the bulk regions has a distinct boundary.}
\label{fig:spatial}
\end{figure}
The brane $\Sigma$ is sandwiched between two $N$-dimensional bulk
spaces $M_1$ and $M_2$, as shown in Fig.~\ref{fig:spatial}.  We
impose $\mathbb{Z}_2$ symmetry across the brane, which implies
that $M_1$ and $M_2$ are ``mirror images'' of one another.  Each
of the bulk spaces is identified with a topological S-AdS${}_N$
manifold. On each side of the brane, we place a bulk coordinate
system $x_\OneTwo = \{ T, \theta^1,\ldots,\theta^d,R_\onetwo \}$
such that the metric $g^\OneTwo_{AB}$ on $M_\onetwo$ is
\begin{equation}\label{bulk metric}
    ds^2_\OneTwo = -F(R_\onetwo)\,dT^2 +
    \frac{dR_\onetwo^2}{F(R_\onetwo)} + R_\onetwo^2
    \sigma^{(k,d)}_{ab} d\theta^a d\theta^b,
\end{equation}
where $A,B = 0 \ldots (d+1)$.  This reduces to the usual
Schwarzschild-AdS line element if we set $k = 1$ and is a solution
of the bulk Einstein field equations
\begin{equation}\label{bulk field eqns}
    G^\OneTwo_{AB} = \Lambda g^\OneTwo_{AB}
\end{equation}
if we set
\begin{equation}\label{F def}
    F(R) = k - \frac{K}{R^{d-1}} + \frac{2\Lambda
    R^2}{d(d+1)}.
\end{equation}
The constant $K$ is linearly proportional to the ADM mass of the
central object in each of the bulk regions:
\begin{equation}\label{K general soln}
    K = \frac{2MG_N}{d-1} \frac{\Omega_d}{\V},
\end{equation}
where $\Omega_d \equiv {\mathcal V}^{(+)}_d$ and we define the
$N$-dimensional Newton's constant by
\begin{equation}\label{kappa}
    \kappa_N^2 = \left( \frac{N-2}{N-3} \right) \Omega_{N-2} G_N
    = \left( \frac{d}{d-1} \right) \Omega_d G_{d+2},
\end{equation}
where $\kappa_N^2$ is the higher-dimensional gravity matter
coupling in Einstein's equations.\footnote{This is a somewhat
different definition than usually found in the literature; i.e.,
$\kappa_N^2 = 8\pi G_N$.  We find (\ref{kappa}) more useful
because it produces the correct law of gravitation in the
Newtonian limit; see refs.~\cite{Sea03a} or \cite{Sea03b} for more
details.} We note that (\ref{bulk metric}) solves (\ref{bulk field
eqns}) even if $K$ is negative. In this paper, we will generally
restrict ourselves to $\Lambda \ge 0$ --- i.e., AdS space ---
although this is not a critical assumption for any of the
derivation.

We should comment on the horizon structure of the bulk manifolds.
In general, there will be a Killing horizon at any $R = R_H$ such
that $F(R_H) = 0$.  For the moment, let us focus in on the most
physically relevant case of $d = 3$.  Then we find the following
solution for $R_H$:
\begin{equation}
R_H^2 =
\begin{cases}
    \sqrt{\frac{6K}{\Lambda}}, & k =0, \\
    - \frac{3k}{\Lambda} \left( -1 \pm \sqrt{1 + \frac{2}{3}
    \Lambda K} \right), & k = \pm 1.
\end{cases}
\end{equation}
For $K > 0$, it is easy to see that there is only one real and
positive solution for $R_H$ for all values of $k$.  When $K < 0$,
there is no horizon for the $k = 0$ and $k = +1$ case.  However,
if $\Lambda |K| < \tfrac{3}{2}$, there will be two positive and
real solutions for $R_H$ in the $k = -1$ case.  As long as
$\Lambda > 0$, $F(a) \rightarrow \infty$ when $a \rightarrow
\infty$.  This prompts us to use the terminology that ``outside
the horizon'' refers to regions with $F(a) > 0$ and that ``inside
the horizon'' refers to regions with $F(a) < 0$. Clearly the
labels are not strictly applicable to the $k = -1$ case and might
not sense when $d > 3$ or $\Lambda = 0$.  However, we do find it
useful to apply these terms to the general situation and with the
preceding caveat we forge ahead.

\begin{figure}
\includegraphics{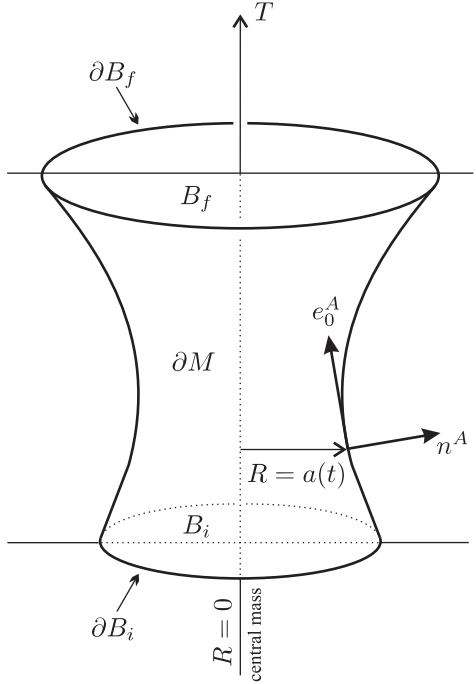}
\caption{A pictorial representation of one of the bulk sections of
our model.  As if Fig.~\ref{fig:spatial}, we suppress $d-1$
dimensions on $\maxsym$ to obtain a 3-dimensional picture.  The
bulk $M$ is the region bounded by $\di M \cup B_i \cup B_f$.}
\label{fig:temporal}
\end{figure}
We now return to the case of arbitrary $d$.  The structure of each
bulk section is shown in Fig.~\ref{fig:temporal}.  The boundary of
$M_\onetwo$ is given by $\di M_\onetwo \cup B_i \cup B_f$.  The
$B_i$ and $B_f$ hypersurfaces are defined by $T = T_i$ and $T_f$
respectively, and will represent the endpoints of temporal
integrations in the action for our model. The other boundary $\di
M_\onetwo$ is described by the hypersurface
\begin{equation}
    R_\onetwo = a(t), \quad T = {T}(t).
\end{equation}
Here, $t$ is a parameter.  In principle, we can identify
$M_\onetwo$ with the portion of the Schwarzschild-AdS manifold
interior or exterior to the $\di M_\onetwo$ world-tube. However,
we want the spatial sections of our model to be compact, so we
take $M_\onetwo$ to be the $(d+2)$-manifold inside $\di
M_\onetwo$. Let us define a function $\Phi(t) \in \mathbb{R}^+$ as
\begin{equation}\label{Phi def}
    \Phi^2 \equiv F(a) \dot{T}^2 - \frac{1}{F(a)} \dot{a}^2,
\end{equation}
where an overdot stands for $d/dt$.  With this definition, we see
that the induced metric on $\di M_1$ is identical to that of $\di
M_2$ and is given by
\begin{equation}\label{brane metric}
    ds_\Sigma^2 = -\Phi^2(t) \, dt^2 + a^2(t) \sigma^{(k,d)}_{ab} d\theta^a
    d\theta^b.
\end{equation}
We identify this with the metric $h_{\alpha\beta}$ on the brane in
the $y = \{ t,\theta^1,\ldots,\theta^d \}$ coordinate system,
where $\alpha,\beta = 0 \ldots d$.  All three of the
$(d+1)$-surfaces $\di M_\onetwo$ and $\Sigma$ are bounded by $\di
B_i$ and $\di B_f$, which can be thought of as $t = t_i$ and $t_f$
$d$-surfaces respectively.\footnote{The boundary times on the
brane are defined by $T_i = T(t_i)$ and $T_f = T(t_f)$.} The
intrinsic geometry of the brane is hence specified by the two
functions $a(t)$ and $\Phi(t)$, which we take to be the brane's
generalized coordinate degrees of freedom. However, it is obvious
that the lapse function $\Phi(t)$ does not represent a genuine
physical degree of freedom because it can be removed from the
discussion via a reparametrization of the brane's time coordinate
$t$.  It is therefore a gauge degree of freedom whose existence
implies that there are first class constraints in our system
\cite{Dir64}. This is important in what follows.

The last ingredient of the model is the matter fields living on
$\Sigma$.  We will characterize the matter degrees of freedom by
the set of ``coordinates'' $\psi = \{ \psi_i \}$.\footnote{Middle
lowercase Latin indices ($i$, $j$, etc.) run over matter
coordinates.} Here, $\psi$ can include scalar or spinorial fields,
perfect fluid velocity potentials \cite{Sch70}, or other types of
fields. At this stage, the only real restriction we place on
$\psi$ comes from the fact that our assumed form of the brane
metric is isotropic and homogeneous, and hence obeys the perfect
cosmological principle (PCP).  This means that $\psi$ can only
depend on $t$ and not $\theta^a$.  We will need the matter
Lagrangian density, which in keeping with the PCP must be of the
form
\begin{equation}
    \Lm = \Lm
    \left( \psi,\dot\psi;a,\Phi \right).
\end{equation}
Notice that $\Lm$ is independent of derivatives of the induced
metric, which is a common and non-restrictive assumption. Later,
we will need the stress energy tensor associated with the matter
fields, which is given by
\begin{equation}\label{stress energy def}
    T_{\alpha\beta} = -2 \frac{ \delta \Lm}{\delta
    h^{\alpha\beta} } + \Lm h_{\alpha\beta}.
\end{equation}

We are now in position to calculate the action of our model.  It
is composed of five parts as follows \cite{Kar02}:
\begin{equation}\label{total action}
    S = \frac{1}{\mathcal N} [ S(M_1) + S(\di M_1) + S(\Sigma) +
    S(\di M_2) + S(M_2) ],
\end{equation}
where ${\mathcal N}$ is a normalization constant that will be
selected later, and
\begin{subequations}
\begin{eqnarray}
    S(M_\onetwo) & = & \int\limits_{M_\onetwo} d^{d+2} x_\OneTwo
   \sqrt{-g^\OneTwo} \, \left[ {\mathcal R}^\OneTwo + 2 \Lambda \right],
   \\ S(\di M_\onetwo) & = & 2 \int\limits_{\di M_\onetwo} d^{d+1}y
   \sqrt{-h} \, \text{Tr} \left[ {\mathcal K}^\OneTwo \right], \\
   S(\Sigma) & = & 2 \kappa^2_N \int\limits_\Sigma d^{d+1}y \sqrt{-h} \,
   \Lm.
\end{eqnarray}
\end{subequations}
In these expressions, ${\mathcal R}^\OneTwo$ is the Ricci scalar
in the bulk regions, $\text{Tr} \left[ {\mathcal K}^\OneTwo
\right]$ is the trace of the extrinsic curvature of $\di
M_\onetwo$.  There is no contribution to the action from the
spacelike boundaries $B_i$ and $B_f$ of $M_\onetwo$ because those
surfaces have vanishing extrinsic curvature.  The $\mathbb{Z}_2$
symmetry immediately gives us
\begin{equation}\label{symmetry 1}
    S(M_1) = S(M_2) \equiv S(M).
\end{equation}
Now, what does this symmetry imply for the boundary terms?  Note
that $\text{Tr} \left[ {\mathcal K}^\OneTwo \right]$ is calculated
with the outward-pointing normal vector field, which means that
the normal on $\di M_1$ is anti-parallel to the normal on $\di
M_2$.  Usually, the $\mathbb{Z}_2$ symmetry gives that the sign of
the extrinsic curvature is inverted as one traverses the brane,
but that assumes a continuous normal vector. In our situation, we
expect $\text{Tr} \left[ {\mathcal K}^{(1)} \right] = \text{Tr}
\left[ {\mathcal K}^{(2)} \right]$, which yields
\begin{equation}\label{symmetry 2}
    S(\di M_1) = S(\di M_2) \equiv S(\di M).
\end{equation}
Hence, we only have to calculate three separate quantities to
arrive at the total action.

The actual calculation of $S$ is straightforward but lengthy,
involving partial integrations, the elimination non-dynamical
contributions, and integrating over the bulk manifolds and spatial
directions on the brane; full details can be found in
refs.~\cite{Sea03a,Sea03b}. The final result is
\begin{equation}\label{the action}
    S = \int_{t_i}^{t_f} dt\, \Phi a^{d-1} \left\{ -
    \frac{\dot{a}}{\Phi}
    \mathrm{arcsinh} \left[ \frac{\dot{a}}{\Phi\sqrt{F(a)}} \right] +
    \sqrt{ \frac{\dot{a}^2}{\Phi^2} + F(a) } + a \alpha_m {\mathcal L}_m
    \right\},
\end{equation}
where we have made the choices
\begin{equation}
    {\mathcal N} \equiv 4d\V, \quad \alpha_m \equiv
    \frac{\kappa_N^2}{2d}.
\end{equation}
This is the effective action for our model.  The degrees of
freedom are the brane radius $a(t)$, the lapse function $\Phi(t)$,
and the matter coordinates $\psi_i(t)$.  We note that under an
arbitrary reparametrization of the time
\begin{equation}
    t \rightarrow \tilde{t} = \tilde{t}(t),
\end{equation}
the lapse transforms as
\begin{equation}
    \Phi \rightarrow \tilde{\Phi} = \Phi \frac{dt}{d\tilde{t}}.
\end{equation}
Assuming that the matter Lagrangian is a proper relativistic
scalar, we see that the total action is invariant under time
transformations.  Systems with this property have Hamiltonian
functions that are formally equal to constraints and hence vanish
on solutions, which is what we will see explicitly in
Sec.~\ref{sec:quantum}. The fact that our action involves
constraints should not be surprising because we have already
identified $\Phi$ as a gauge degree of freedom.  Also, the zero
Hamiltonian phenomena is a trademark of fully covariant theory
such as general relativity.

Before leaving this section, we note that our action can become
imaginary for certain values of $(a,\dot{a})$; i.e., when $F(a) <
0$.  To get around a complex action, we consider the following
identities:
\begin{subequations}
\begin{eqnarray}\label{first identity}
    \ln(i) & = & \mathrm{arcsinh}(iz) - \mathrm{arccosh}(z), \\
    \label{second identity} \ln(-1) & = & \mathrm{arccosh}(z) + \mathrm{arccosh}(-z).
\end{eqnarray}
\end{subequations}
There are a couple of subtle points that one must remember when
working with these complex identities.  The first is that we must
specify a branch cut in order to evaluate the logarithms of
complex quantities.  In all cases, we assume that the argument of
complex numbers lies in the interval $(-\pi,\pi]$ so that $\ln(i)
= i\pi/2$ and $\ln(-1) = i\pi$.  This cut also makes the square
root function single-valued on the negative real axis; we have
that $\sqrt{-x} = i \sqrt{x}$ for all $x \in \mathbb{R}^+$. The
other issue is the fact that the arccosh function is multi-valued
on the positive real axis.  We will always take the principle
branch, with $x \in [1,\infty)$ implying that $\mathrm{arccosh}(x)
> 0$.

\begin{figure}
    \includegraphics{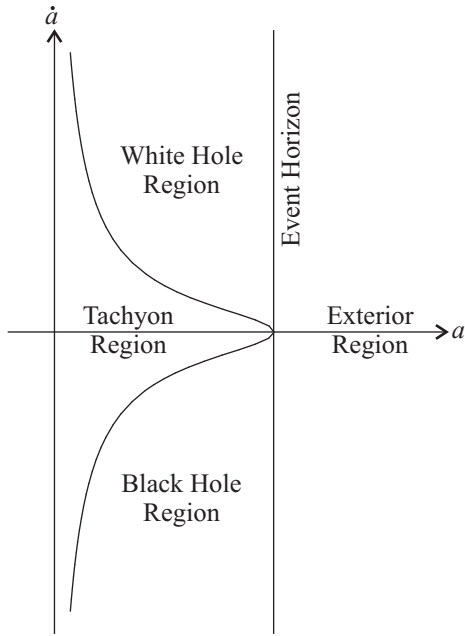}
    \caption{A sketch of the  various regions in the position-velocity
    $(a,\dot{a})$-plane describing our model when the bulk manifold
    contains an event (Killing) horizon.}\label{fig:structure}
\end{figure}

Having clarified our choices for the structure of the complex
plane, let us now define a pair of alternative actions $S_\pm$ by
\begin{equation}\label{alt action}
    S_\pm = \int_{t_i}^{t_f} dt\, \Phi a^{d-1} \left\{ \mp
    \frac{\dot{a}}{\Phi} \mathrm{arccosh} \left[
    \frac{\pm\dot{a}}{\Phi\sqrt{-F(a)}} \right] + \sqrt{
    \frac{\dot{a}^2}{\Phi^2} + F(a) } + a \alpha_m {\mathcal L}_m
    \right\}.
\end{equation}
Using the identity (\ref{first identity}) it is not difficult to
show that
\begin{equation}
    S - S_+ = -\frac{ \ln (i) }{n} \int_{t_i}^{t_f} dt \frac{d}{dt} a^d,
\end{equation}
and using (\ref{second identity}) we have
\begin{equation}
    S_+ - S_- = -\frac{ \ln (-1) }{n} \int_{t_i}^{t_f} dt \frac{d}{dt}
    a^d.
\end{equation}
Therefore, the three actions $S$ and $S_\pm$ differ by terms
proportional to the integral of total time derivatives.  This
means that the variations of each are the same, and each one is a
valid action for our model.  Now, each action will be real and
well-behaved for different regions of the $(a,\dot{a})$-plane,
which are depicted in Fig.~\ref{fig:structure} and defined by
\begin{subequations}
\begin{eqnarray}
    \text{Exterior Region} & \equiv & \bigg\{
    a \in \mathbb{R}^+,\, \dot{a} \in \mathbb{R}
    \, \bigg| \, F(a)
    > 0 \bigg\}, \\
    \text{White Hole Region} & \equiv & \left\{ a \in \mathbb{R}^+,\,
    \dot{a} \in \mathbb{R} \, \bigg| \, F(a)
    < 0, \, 0 < \dot{a}, \, 0 <  \frac{\dot{a}^2}{\Phi^2}
    + F(a) \right\}, \\
    \text{Black Hole Region} & \equiv & \left\{ a \in \mathbb{R}^+,\,
    \dot{a} \in \mathbb{R} \, \bigg| \, F(a)
    < 0, \, \dot{a} < 0, \, 0 <  \frac{\dot{a}^2}{\Phi^2}
    + F(a) \right\}, \\
    \text{Tachyon Region} & \equiv & \left\{ a \in \mathbb{R}^+,\,
    \dot{a} \in \mathbb{R} \, \bigg| \, \frac{\dot{a}^2}{\Phi^2}
    + F(a) < 0 \right\}.
\end{eqnarray}
\end{subequations}
The original action $S$ is real-valued in the exterior region,
while the alternative actions $S_\pm$ are well-behaved in the
white and black hole regions respectively.  The adjectives ``White
Hole'' and ``Black Hole'' are used because the brane is moving
away from the central singularity when it is in the former region
and towards the singularity when in the latter.  The fourth region
is intriguing; when the brane is inside it all of the actions we
have written down thus far fail to be real.  Also, when inside
this region it is impossible to solve equation (\ref{Phi def}) for
$\Phi \in \mathbb{R}$, which means the brane's ``timelike''
tangent vector $e^A_0$ becomes spacelike. Since the brane behaves
like a particle with imaginary mass in this portion of the
position-velocity plane, we label it as the ``Tachyon Region''.
What is the form of the action for our model inside the tachyon
region?  To answer this, we can analytically continue the $S_\pm$
actions by using the identities
\begin{subequations}
\begin{eqnarray}
    \label{arccosh identity} 0 & = & \text{arccosh}(z) - i \, \mathrm{arccos}(z), \\
    \pi & = & \text{arccos}(x) + \text{arccos}(-x).
\end{eqnarray}
\end{subequations}
When the first of these is applied to the $S_\pm$ actions, we
obtain two distinct expressions.  But if we apply the second
identity to the action derived from $S_-$ and discard a boundary
term, we arrive at the tachyon action:
\begin{equation}\label{tachyon action}
    S_\mathrm{tach} = i\int_{t_i}^{t_f} dt\, \Phi a^{d-1} \left\{
    - \frac{\dot{a}}{\Phi}
    \mathrm{arccos} \left[ \frac{\dot{a}}{\Phi\sqrt{-F(a)}} \right]
    + \sqrt{ -\left[ \frac{\dot{a}^2}{\Phi^2} + F(a) \right] } - ia \alpha_m {\mathcal L}_m
    \right\}.
\end{equation}
This action is valid throughout the tachyon region, and is
explicitly complex-valued.  Recall that when the action for a
mechanical system is complex when evaluated along a given
trajectory that solves the equation of motion, that trajectory is
considered to be classically forbidden. In our case, this means
that the tachyon region is inaccessible by the brane in the
context of classical mechanics.  We will revisit this notion in
the context of semi-classical considerations shortly.

To summarize, we have obtained the reduced action(s) governing the
motion of the brane in our model.  The degrees of freedom are the
brane radius, the lapse function, and whatever coordinates we need
to describe the matter living on the brane.  When there are
horizons in the bulk, the $(a,\dot{a})$-plane acquires the
structure depicted in Fig.~\ref{fig:structure}.  We find that four
actions are needed in this case: $S$, $S_\pm$, and
$S_\mathrm{tach}$. These actions differ by integrals of time
derivatives, and are hence equivalent.

\section{The dynamics of the classical cosmology}\label{sec:classical}

\subsection{The Friedman equation exterior to the tachyon region}
\label{sec:Friedman}

We now turn our attention to the classical dynamics of our system.
The equation of motion that we will be primarily concerned with
can be obtained from varying the action with respect to the lapse
$\Phi$ and setting the result equal to zero. But before we do
this, recall that we derived four distinct expressions for the
action in the previous section. This might cause one to wonder:
which action must we vary in order to obtain the correct equation
of motion? The answer is that it does not matter, each of the
actions $S$, $S_\pm$ and $S_\mathrm{tach}$ differ from one another
by boundary contributions and must therefore yield the same
equations of motion.  We can explicitly confirm this by
calculating the functional derivatives
\begin{equation}\label{dS/dPhi}
    \frac{\delta S}{\delta \Phi} = \frac{\delta S_\pm}{\delta \Phi}
    = \frac{\delta S_\mathrm{tach}}{\delta \Phi}
    = \int dt \, a^{d-1} \, \left[
    \sqrt{ \frac{\dot{a}^2}{\Phi^2} + F } + a\alpha_m \left(
    \Lm + \Phi \frac{ \di \Lm }{ \di \Phi} \right) \right] \stackrel
    {{\mathrm{set}}}{=} 0.
\end{equation}
Note that we have evaluated each derivative in the region where
the associated action is valid; i.e., $\delta S / \delta \Phi$ is
evaluated in the exterior region, $\delta S_\mathrm{tach} / \delta
\Phi$ is evaluated in the tachyon region, and so on.\footnote{In
the interests of concise notation, we omit the limits of
integration and functional dependence of $F$ on $a$ from this and
subsequent formulae.}  We see that all four actions yield the same
equation of motion, namely
\begin{equation}\label{primative EOM}
    0 = \sqrt{ \frac{\dot{a}^2}{\Phi^2} + F } + a\alpha_m \left(
    \Lm + \Phi \frac{ \di \Lm }{ \di \Phi} \right).
\end{equation}
To simplify this, recall our formula for the stress energy tensor
on the brane (\ref{stress energy def}), which implies
\begin{equation}
    T_{00} = -2 \frac{\di \Lm}{\di h^{00}} + \Lm h_{00}.
\end{equation}
But, we have $h_{00} = -\Phi^2$ and $h^{00} = -\Phi^{-2}$.  This
results in
\begin{equation}\label{a}
    \Lm + \Phi \frac{ \di \Lm }{ \di
    \Phi} = -\frac{T_{00}}{\Phi^2}.
\end{equation}
Now, consider the total density of matter on the brane as measured
by comoving observers $\rho_\mathrm{tot}$.  These observers have
$(d+1)$-velocities $u_\alpha = -\Phi \, \di_\alpha t$, so the
measured density is
\begin{equation}\label{b}
    \rho_\mathrm{tot} = u^\alpha u^\beta T_{\alpha\beta} =
    \frac{T_{00}}{\Phi^2}.
\end{equation}
Putting (\ref{a}) and (\ref{b}) into (\ref{primative EOM}) yields
\begin{equation}\label{Lagrange density}
    \Lm + \Phi \frac{ \di \Lm }{ \di
    \Phi} = -\rho_\mathrm{tot}
\end{equation}
and
\begin{equation}\label{classical EOM}
    0 = \sqrt{ \frac{\dot{a}^2}{\Phi^2} + F } - a \alpha_m
    \rho_\mathrm{tot}.
\end{equation}
It should be stressed that equation (\ref{Lagrange density}) is
quite general and not limited to the perfect fluid case, which is
the prime example that we consider below.  Notice that if
$\rho_\mathrm{tot}$ is taken to be real, then the equation of
motion implies that
\begin{equation}\label{tachyon inequality}
    0 < \frac{\dot{a}^2}{\Phi^2} + F.
\end{equation}
This confirms that the tachyon portion of the $(a,\dot{a})$-plane
is classically forbidden.

Equation (\ref{classical EOM}) can be rewritten as a sort of
energy conservation equation
\begin{subequations}\label{energy conservation}
\begin{eqnarray}
    0 & = & \tfrac{1}{2}\dot{a}^2 + V, \\
    V & \equiv & \tfrac{1}{2} \Phi^2 (F- a^2 \alpha_m^2 \rho_\mathrm{tot}^2),
\end{eqnarray}
\end{subequations}
or in an explicitly Friedman-like form
\begin{equation}\label{friedmann}
    \frac{\dot{a}^2}{a^2} = \Phi^2 \left[ \alpha_m^2
    \rho_\mathrm{tot}^2 +\frac{K}{a^{d+1}}
    - \frac{2\Lambda}{d(d+1)} - \frac{k}{a^2} \right].
\end{equation}
Each form of the classical equation of motion is useful in
different contexts.  In these equations, we still retain $\Phi$ as
a gauge degree of freedom that can be specified arbitrarily. Two
special choices of gauge are
\begin{equation}
    \begin{array}{cclcl}
        \left\{ \Phi = 1, t = \tau \right\} & \,\, \Rightarrow \,\, & ds^2_\Sigma =
        -d\tau^2 + a^2(\tau) \, d\sigma_{(k,d)}^2 & \,\, & \text{(proper
        time gauge)}, \\ \left\{ \Phi = a, t = \eta \right\} & \,\, \Rightarrow \,\,
        & ds^2_\Sigma = a^2(\eta)\left[-d\eta^2 + d\sigma_{(k,d)}^2\right]
        & & \text{(conformal time gauge)}.
    \end{array}
\end{equation}

Let us now concentrate on the energy conservation equation
(\ref{energy conservation}).  This formula allows us to make an
analogy between the brane's radius $a(t)$ and the trajectory of a
zero-energy particle moving in a potential $V$.  At a classical
level, such a particle cannot exist in regions where the potential
is positive.  This fact allows us to identify brane radii which
are classically allowed and classically forbidden:
\begin{equation}\label{classically allowed/forbidden}
    \begin{array}{rclcl}
        F(a) & > & a^2 \alpha_m^2 \rho_\mathrm{tot}^2 & \,\,\Rightarrow\,\, &
        \text{classically forbidden}, \\
        F(a) & < & a^2 \alpha_m^2 \rho_\mathrm{tot}^2 & \,\,\Rightarrow\,\, &
        \text{classically allowed}.
    \end{array}
\end{equation}
We should make it clear that classically forbidden regions defined
in this way are distinct from the previously discussed tachyon
region.  It is interesting to note that the black and white hole
regions of configuration space have $F < 0$ by definition;
therefore, each region is always classically allowed.

Now, the existence of classically forbidden regions exterior to
the tachyon sector raises the possibility that $a(t)$ may be
bounded from below, above, or above and below; these possibilities
imply that the cosmology living on the brane may feature a big
bounce, a big crunch, or oscillatory behaviour respectively.  The
existence of barriers in the cosmological potential also allows
for the quantum tunnelling of the universe between various
classically allowed regions.  But before we get too far ahead of
ourselves, we note that without specifying the matter fields on
$\Sigma$, it is impossible to know if classical forbidden regions
exist or not. In Sec.~\ref{sec:special case}, we will study a
special case in some detail to see under which circumstances
potential barriers manifest themselves.

Before leaving this section, we attempt to gain some intuition
about the physics of the Friedmann equation by studying the
Newtonian limit of (\ref{classical EOM}).  Let us momentarily
limit the discussion to the $k = 1$ case in the proper time gauge,
and expand (\ref{classical EOM}) in the limit of small velocities
$\dot{a} \ll 1$, small mass $K \ll a^{d-1}$, and vanishing vacuum
energy $\Lambda = 0$. Using formulae (\ref{K general soln}) and
(\ref{kappa}) with $k = 1$, we find
\begin{equation}
    0 = m + \frac{1}{2}m\dot{a}^2 - \frac{G_N Mm}{(d-1) a^{d-1}}
    - \frac{1}{2} \frac{G_N m^2}{(d-1) a^{d-1}},
\end{equation}
where $m = \rho_\mathrm{tot} \Omega_d a^d$ is the mass of the
matter on the brane, and $M$ is the ADM mass of the black hole. On
the righthand side, the first term is the brane's rest mass
energy, the second term is its kinetic energy, the third is the
gravitational potential energy due to the black hole, and the
fourth is the gravitational self-energy of $\Sigma$, which can be
thought of as a massive $d$-spherical shell.\footnote{The $(d-1)$
factors appear so that when potential energies are differentiated,
they yield the correct force laws.} Therefore, on a Newtonian
level, the $k=1$ brane behaves as a massive spherical shell
surrounding a central body with zero-total energy.  We mentioned
above that situations with classically forbidden regions are of
special interest.  For this situation, the brane can obviously
never achieve infinite radius, so there is at least one
classically forbidden region $(a_\text{max},\infty)$.  We can
engineer another forbidden region if we allow the black hole mass
to become negative.  In such a situation, the dynamics is
dominated by the competition between the tendency of a
self-gravitating shell to collapse on itself and the repulsive
nature of the central object.  We will see a fully-relativistic
example of this effect in the next section.

\subsection{Exact analysis of a special case}\label{sec:special
case}

In this section, we will concentrate on a special case of the
classical cosmology that allows for some level of exact analysis.
We will make some arbitrary parameter choices that are not meant
to convey some sort of advocacy, we are merely attempting to write
down a model that is easy to deal with mathematically. First, we
assume
\begin{equation}
    \{ d = 3, \Lambda = 0, k = 0, \Phi = H_0^{-1} \}.
\end{equation}
In other words, we identify $\Sigma$ with a spatially flat
$(3+1)$-dimensional FLRW universe, tune the bulk cosmological
constant to zero, and set the lapse function equal to the current
value of the Hubble parameter, defined as
\begin{equation}
    H = \frac{1}{a} \frac{da}{d\tau} = H_0 \frac{\dot{a}}{a},
\end{equation}
where $\tau$ is the proper cosmic time.  (We use the term
``current'' to refer to the epoch with $a = 1$.)  This gives the
initial condition
\begin{equation}
    a = 1 \quad \Rightarrow \quad \dot{a} = 1.
\end{equation}
Essentially, all we have done is identify $t$ with the
dimensionless Hubble time to simplify what follows.  For the
matter fields, we take
\begin{equation}
    {\mathcal L}_m =  {\mathcal L}_\mathrm{v} + {\mathcal L}_\mathrm{d}.
\end{equation}
Here, ${\mathcal L}_\mathrm{v}$ is the Lagrangian density of
perfect fluid matter with a vacuum-like equation of state
$\rho_\mathrm{v} = -p_\mathrm{v}$ while ${\mathcal L}_\mathrm{d}$
corresponds to dust-like matter with equation of state
$p_\mathrm{d} = 0$.\footnote{See Appendix \ref{app:dust} for the
definition and discussion of perfect fluid Lagrangian densities.}
The total density of matter on the brane is then given by
\begin{equation}
    \rho_\mathrm{tot} = \rho_\mathrm{v} + \rho_\mathrm{d} a^{-3},
\end{equation}
where we have made use of equation (\ref{rho soln}).  The Friedman
equation in this case can therefore be written as
\begin{equation}\label{reduced friedmann}
    \frac{\dot{a}^2}{a^2} = \Omega_x a^{-3} + \Omega_y a^{-6} +
    \Omega_z + \Omega_w a^{-4},
\end{equation}
where the (current epoch) density parameters are defined as
\begin{equation}
    \Omega_x \equiv
    \frac{2\alpha_m^2\rho_\mathrm{d}\rho_\mathrm{v}}{H_0^2}, \,
    \Omega_y \equiv \frac{\alpha_m^2 \rho_\mathrm{d}^2}{H_0^2}, \,
    \Omega_z = \frac{\alpha_m^2 \rho_\mathrm{v}^2}{H_0^2}, \,
    \Omega_w = \frac{K}{H_0^2}.
\end{equation}
This parameters are not freely specifiable, they are constrained
by
\begin{subequations}
\begin{eqnarray}
    0 & = & \Omega_x + \Omega_y + \Omega_z + \Omega_w - 1, \\
    0 & = & \Omega_x^2 - 4\Omega_y \Omega_z.
\end{eqnarray}
\end{subequations}
This means that there are effectively two free parameters in
(\ref{reduced friedmann}).  If we take the two independent
parameters to be $\Omega_x$ and $\Omega_y$, each solution of the
Friedman equation in this situation corresponds to a point in the
parameter space ${\mathcal P}^2 = (\Omega_x,\Omega_y)$.

We can interpret (\ref{reduced friedmann}) as implying that the
cosmological dynamics on the brane are driven by four
constituents.  The $\Omega_x$ parameter refers to a cosmological
dust population, and $\Omega_y$ corresponds to matter whose
density depends quadratically on the dust.  The vacuum energy on
the brane is characterized by $\Omega_z$.  Finally, $\Omega_w$
seems to be associated with some radiation field whose amplitude
is linearly related to the mass of the higher-dimensional black
hole.  In the standard brane world lexicon, the $w$ field is
called Weyl or ``dark'' radiation.  It reflects the contribution
of the higher dimensional Weyl tensor to the intrinsic geometry of
the brane, and its appearance in this context is hardly
surprising.

\begin{figure*}
\includegraphics{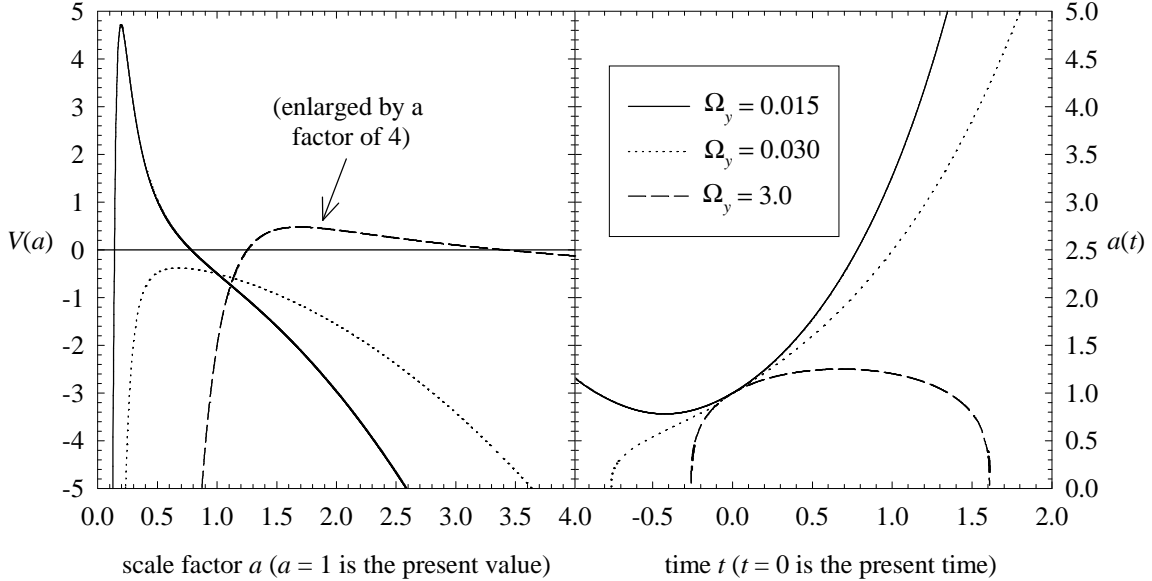}
\caption{The cosmological potential $V(a)$ for the special case
discussed in Sec.~\ref{sec:special case} (\emph{left}) along with
the associated numerical solutions (\emph{right}). We have taken
$\Omega_x = 0.3$. As indicated in the left panel, the $\Omega_y =
3.0$ potential curve has been scaled by a factor of 4 to highlight
the existence of the (shallow) barrier.  The numeric solutions are
calibrated so that the scale factor at the current time is
unity.}\label{fig:classical}
\end{figure*}

The cosmological potential that appears in equation (\ref{energy
conservation}) for this case is
\begin{equation}\label{flat f}
    V(a) = - \left[ \frac{(\Omega_x a^3 + 2 \Omega_y)^2 + 4 \Omega_y
    \Omega_w a^2}{8\Omega_y a^4} \right] ,
\end{equation}
where
\begin{equation}
    \Omega_w = 1 - \frac{(\Omega_x + 2 \Omega_y)^2}{4\Omega_y}.
\end{equation}
Some obvious properties of the potential are
\begin{equation}\label{f properties}
    V(1) = -\tfrac{1}{2}, \quad \lim_{a \rightarrow 0} V(a) =
    \lim_{a \rightarrow \infty} V(a) = -\infty.
\end{equation}
By definition, $\Omega_y$ is positive definite so the potential
will be negative definite if $\Omega_w > 0$; i.e., if $K > 0$.
Therefore, if the bulk black hole mass is positive, there are no
classically forbidden regions in this special case.  This
conclusion matches the Newtonian conclusion of the previous
section.  We can see explicitly forbidden regions by plotting the
potential for some particular values of $(\Omega_x,\Omega_y)$
along with numeric scale factor solutions, which is done in
Fig.~\ref{fig:classical}. Notice that when we solve the Friedman
equation numerically in the context of our special case, we are
obliged to use the initial condition $a = 1$ at the current epoch,
which we define to be at $t = 0$.  Two of the situations in
Fig.~\ref{fig:classical} show classically forbidden regions, which
manifest themselves as big bounce/crunch cosmologies with
$\Omega_w < 0$.

\begin{figure}[t]
\begin{center}
\includegraphics{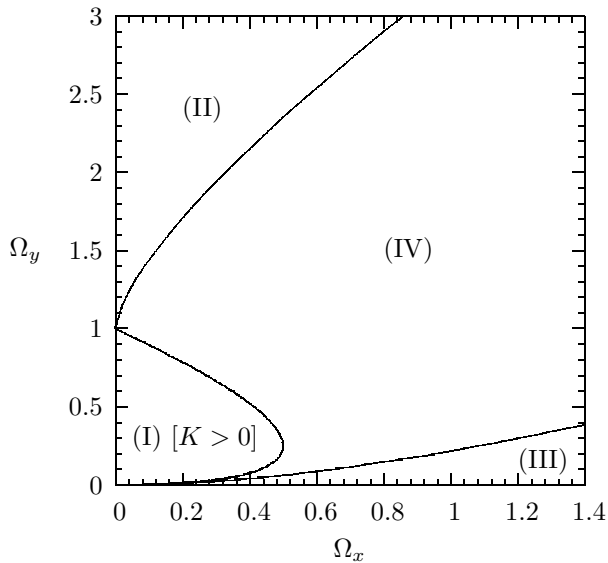}
\end{center}
\caption[Structure of brane cosmology parameter space]{The
$\mathcal{P}^2 = (\Omega_1,\Omega_2)$ parameter space for the
special brane cosmology discussed in Section \ref{sec:special
case}.  The four regions (I)--(IV) are defined in Table
\ref{7:tab:logic}; for example, if $(\Omega_1,\Omega_2)$ lies
within region (III) then we know that we have a big bounce
cosmology.  Notice that the only region where the bulk black holes
have positive mass is region (I).\label{7:fig:parameter}}
\end{figure}

We conclude the present analysis by presenting analytic
expressions that allow one to predict the qualitative behaviour of
the scale factor solutions given the values of
$(\Omega_x,\Omega_y)$.  This problem reduces to characterizing the
behaviour of a cubic polynomial and is presented in detail
elsewhere \cite{Sea03a,Sea03b}; here, we merely present the
results. Consider the following three conditions:
\begin{subequations}\label{inequalities}
\begin{eqnarray}\label{weyl inequality}
    h_1: & & 4\Omega_y - (\Omega_x + 2\Omega_y)^2 > 0, \\ \label{discriminant cond}
    h_2: & & \left[ (\Omega_x + 2 \Omega_y)^2 - 4\Omega_y \right]^{3/2} -
    27 \Omega_x \Omega_y^2 > 0, \\ \label{big root cond}
    h_3: & & (\Omega_x + 2 \Omega_y)^2 - 4\Omega_y - 9\Omega_x^2 >
    0.
\end{eqnarray}
\end{subequations}
The origin and fate of the brane universe depends on whether these
hypotheses are true or false, as communicated in Table
\ref{7:tab:logic}. We plot these inequalities in the
$\mathcal{P}^2$ parameter space in Figure \ref{7:fig:parameter}.
\begin{table*}
\begin{center}
\begin{small}
\begin{tabular}{|c|c||c|c|}
\hline \multicolumn{2}{|c||}{\textit{Conditions}} & $h_3$ true &
$h_3$ false \\
\hline \hline \multicolumn{2}{|c||}{$h_1$ true} &
\multicolumn{2}{|c|} {big bang \& eternal expansion (I) } \\
\hline $h_1$ false & $h_2$ true &
\begin{minipage}[t]{1.5in}
\begin{center}
    big bang \& crunch (II)
\end{center}
\end{minipage} &
\begin{minipage}[t]{1.5in}
\begin{center}
    big bounce (III)
\end{center}
\end{minipage} \\ \cline{2-4}
 & $h_2$ false & \multicolumn{2}{|c|}{big bang \&
eternal expansion (IV)} \\ \hline
\end{tabular}
\end{small}
\caption[Qualitative behaviour of brane cosmologies]{The
qualitative behaviour of the special brane cosmology discussed in
Section \ref{sec:special case}.  The origin and fate of the
universe is determined by whether the $\{h_1,h_2,h_3\}$ hypotheses
are true or false.}\label{7:tab:logic}
\end{center}
\end{table*}

To summarize, in this subsection we have examined the classical
cosmology of a special case of our model.  The case considered is
characterized by $d = 3$, a vanishing bulk cosmological constant,
and spatially flat submanifolds $\maxsym = S_3^{(0)}$. We took the
matter on the brane to be given by a dust population plus a vacuum
energy contribution.  We found that the potential governing the
brane's motion could have classically forbidden regions if the
bulk black hole has negative mass.  Finally, we showed that the
parameter space labelling solutions of the Friedmann equation was
2-dimensional, and we analytically determined the origin and fate
of the universe on the brane based on the values of those
parameters.

We finish by noting that despite the fact that the preceding
scenario seems somewhat contrived, it is not wholly unphysical. We
do indeed live in a 4-dimensional universe that appears to be
spatially flat and contains a vacuum energy and cosmological dust.
Indeed, from the point of view of observational cosmology, a
simplistic model for our universe could be realized by setting the
dust density parameter $\Omega_x = 0.3$ and the amplitude of the
quadratic correction to be $\Omega_y \ll 1$, which is necessary to
avoid messing up nucleosynthesis. The dark matter in such a model
comes from the Weyl contribution $\Omega_w$.  The only thing that
seems strange is our allowance for negative mass in the bulk. We
will not attempt to argue that this is or is not reasonable, other
than to reflect on fact that in order to realize classically
forbidden regions in standard cosmology, one often needs to break
the energy conditions.  At least in this special case, this truism
carries over to the brane world scenario.

\subsection{Instanton trajectories and tachyonic
branes}\label{sec:instanton}

We conclude our analysis of the classical mechanics of our model
by wandering into the semi-classical regime and considering brane
instanton trajectories.  We are especially interested in showing
how the presence of the bulk black holes alters the archetypical
example of the quantum birth of the universe: namely the deSitter
FLRW model with spherical spatial sections treated in the
semi-classical approximation.  We also consider classical
trajectories that traverse the tachyon region, and find that such
paths can only be realized if the branes density is allowed to
become imaginary.

\begin{figure*}
\begin{center}
\includegraphics{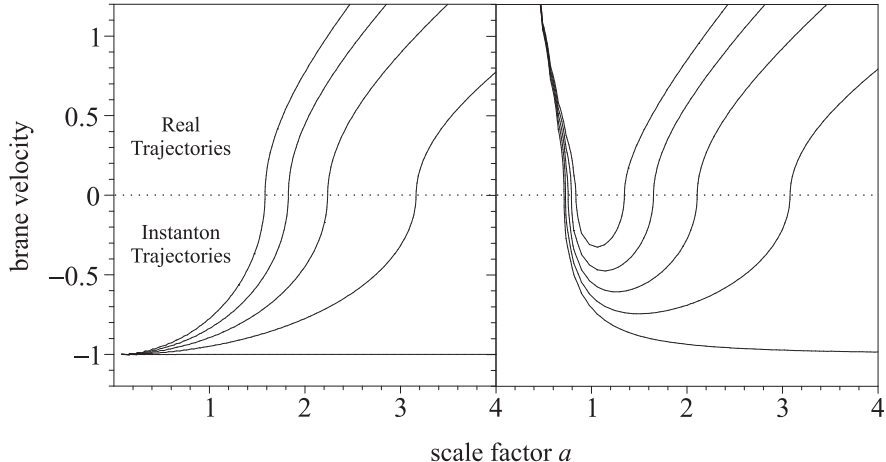}
\caption{The trajectory of spherical 3-branes through
configuration space for the case where $\Phi = 1$ and the brane
contains only vacuum energy.  Different curves correspond to
$\alpha_m^2 \rho_0^2 = 0,0.1,0.2,0.3,0.4$ from bottom to top.  The
left panel shows the canonical deSitter instanton with $K =
\Lambda = 0$ while the right panel shows how the trajectories are
deformed when $K = \tfrac{1}{2}$.\label{fig:instanton}}
\end{center}
\end{figure*}

For this section, let us also choose $d=3$, set the lapse equal to
unity, and the matter content of the brane to be that of a single
perfect fluid with equation of state $p = \gamma \rho$.  Then the
classical brane trajectory can be written as
\begin{equation}
    \alpha_m^2 \rho_0^2 = a^{2(3\gamma +2)} \left[ \left( \frac{da}{dt} \right)^2 +
    F(a) \right].
\end{equation}
Here, $\rho_0$ is a constant that controls the amplitude of the
matter field.  Instanton trajectories can be found from this by
making the switch to the Euclidean time $t \rightarrow i \tau_E$.
The ordinary $k = 1$ deSitter model is found by setting $K =
\Lambda = 0$ and $\gamma = -1$.  To construct the trajectory of a
universe that is ``created from nothing'', we replace the
classical trajectory with the instanton trajectory whenever $da/dt
< 0$.  The brane's path through configuration space for this setup
is shown in the left panel of Fig.~\ref{fig:instanton} for several
different values of $\alpha_m^2 \rho_0^2$ ranging from $0$ to
$0.4$. In this plot we see the familiar behaviour of the deSitter
instanton; all of the Euclidean trajectories interpolate between
the ordinary expanding universe and a universe of zero radius,
which is the ``nothing'' state.  Now what happens if we turn on
the bulk black hole mass?  We set $K = \tfrac{1}{2}$ and replot
the trajectories for the same choices of $\alpha_m^2 \rho_0^2$ in
the right panel of Fig.~\ref{fig:instanton}.  Now the instanton
trajectories interpolate between an expanding state with a
conventional big bang and an eternally expanding universe.
Essentially, the black holes create a classically allowed region
around the singularity that is not present in the archetypical
case. Physically, one can understand this by realizing that the
gravitational attraction of the black hole is in direct
competition with the tendency of a spherical shell of vacuum
energy to inflate.  The important thing is that the black holes
essentially expels the instanton paths from the $a = 0$ area,
breaking up the creation from nothing picture.

\begin{figure*}
\begin{center}
\includegraphics{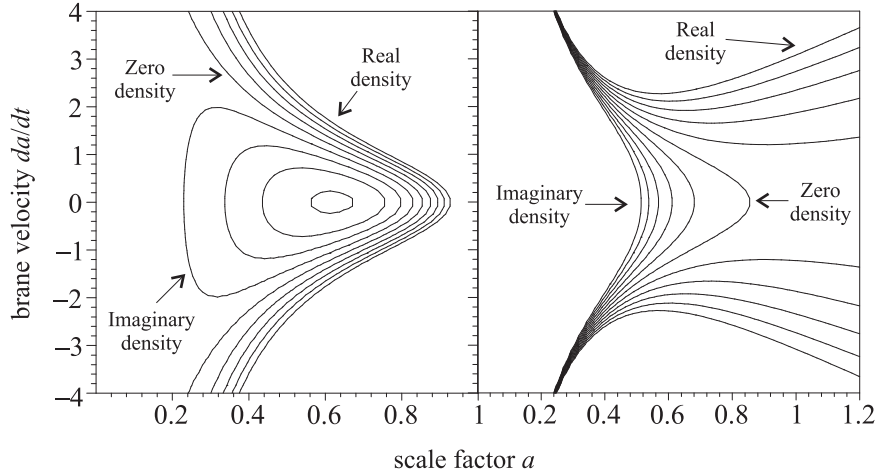}
\caption{The trajectory of spherical 3-branes containing real and
tachyonic matter through configuration space.  We set $\Phi = 1$,
$K =1$ and $\Lambda = \tfrac{1}{2}$.  The left panel shows dust
filled branes with $\alpha_m^2 \rho_0^2$ ranging from $-0.2$ to
$+0.2$.  The right panel shows vacuum dominated branes with
$\alpha_m^2 \rho_0^2$ ranging from $-10$ to
$+10$.\label{fig:contours}}
\end{center}
\end{figure*}

It is interesting to note that in Fig.~\ref{fig:instanton} that
the instanton trajectories do not seem to intersect the tachyon
region.  One can confirm that this is true in all situations by
applying the Wick rotation of the time to inequality (\ref{tachyon
inequality}):
\begin{equation}
    F(a) > \frac{1}{\Phi^2} \left( \frac{da}{d\tau_E} \right)^2.
\end{equation}
So, we see that the Euclidean trajectory only exists for $F(a)
> 0$; i.e., in the exterior region.  This is strange because we
usually associate instanton trajectories with all of the
classically forbidden regions of a model; clearly, the tachyon
region is a special kind of forbidden region and actually
represents an insurmountable boundary at the semi-classical level.
Again this makes sense when we realize that the brane would have
to become spacelike if it entering the tachyon region; it seems as
if there is no semi-classical amplitude for such a transition.
This also suggests how we might be able to find trajectories
inside the region.  Suppose that the brane is populated by
tachyonic matter; i.e., matter with imaginary density $\alpha_m^2
\rho^2_0 < 0$. In that case we find that
\begin{equation}
    0 > \frac{\dot{a}^2}{\Phi^2} + F(a),
\end{equation}
which means that the trajectory is entirely contained within the
tachyon region.  We plot some of the configuration space
trajectories of branes containing real and tachyonic dust and
vacuum matter in Fig.~\ref{fig:contours}.  Again we set $d = 3$
and $\Phi = 1$, we also choose $K =1$ and $\Lambda =
\tfrac{1}{2}$.  While the branes with real dust go through a big
bang and big crunch, the branes with tachyonic dust are seen to go
through periodic expansion and contraction. In contrast, branes
with real vacuum energy begin with a big bang and expand for ever
(or vice versa) while branes with imaginary vacuum energy go
through a big bang and big crunch.

In conclusion, in this section we have considered the instanton
trajectories of spherical vacuum 3-branes, some of which are
plotted in Fig.~\ref{fig:instanton}. We have seen how the presence
of the bulk black holes ruins the ``creation from nothing''
scenario associated with the purely 4-dimensional FLRW model.  We
have also seen that the only way to find brane trajectories that
pass through the tachyon region is to have tachyonic matter on the
brane.  Such paths are shown with their conventional counterparts
in Fig.~\ref{fig:contours}.

\section{Hamiltonian formalism and the quantum
cosmology}\label{sec:quantum}

We now want to pass over from the variational formalism used up to
this point to the Hamiltonian structure needed to quantize our
model.  But we immediately run into an ambiguity due to the fact
that we have at least three different actions that we can use to
Hamiltonize the model.\footnote{We really do not need to worry
about that tachyon action $S_\mathrm{tach}$ since it is the simple
analytic continuation of $S_+$, so it is sufficient to consider
the latter only.}  The simplest thing to do is just choose one
action --- $S$ say --- and ignore the others.  But the fact that
$S$ is complex in the interior region means that the momenta
derived from $S$ are complex there too.  So if we decide to use
$S$ to describe the dynamics of the brane throughout configuration
space we are forced to deal with a complex phase space.  There is
a similar problem with using either $S_\pm$ as the exclusive
action for the model. While the issue of complex phase space in
and of itself is intriguing, we are not looking for that level of
complication in the current study.  So we pursue an alternative
line of attack: we simply take the action of the model to be
defined in a piecewise fashion over configuration space; i.e., we
take $S$ in the exterior region and $S_\pm$ in the white/black
hole regions.\footnote{The adoption of a piecewise action is not
unique to this study; Corchi \emph{et al.}~\cite{Cor01} used a
similar procedure when considering the quantum collapse of a small
dust shell.} The hope is that the Wheeler-DeWitt equation arising
from each of the actions will match smoothly across the boundary
between these regions. After a suitable manipulation of the
constraints, we will see that this hope can be borne out.  We
first carry out the Hamiltonization in the exterior region, and
then do the same for the interior regions
--- the procedures are virtually identical.

\subsection{The exterior region}\label{sec:negative mass}

In the exterior region, we can describe our system by one action
$S = \int dt\,L$, given by equation (\ref{the action}), and we may
assume $\sqrt{F} \in \mathbb{R}^+$. We define the model's
Lagrangian by
\begin{subequations}
\begin{eqnarray}
    L & = & L_g + \alpha_m L_m, \\
    L_g & = & \Phi a^{d-1} \left[ -
    \frac{\dot{a}}{\Phi}
    \mathrm{arcsinh} \left( \frac{\dot{a}}{\Phi\sqrt{F}} \right) +
    \sqrt{ \frac{\dot{a}^2}{\Phi^2} + F } \right], \\ L_m & =
    & \Phi a^d {\mathcal L}_m.
\end{eqnarray}
\end{subequations}
The canonical momenta conjugate to the scale factor and lapse
function are simply found:
\begin{equation}
    p_a \equiv \frac{\di L}{\di \dot{a}} = -a^{d-1}
    \mathrm{arcsinh} \left( \frac{\dot{a}}{\Phi\sqrt{F}} \right),
    \quad p_\Phi \equiv \frac{\di L}{\di \dot{\Phi}} = 0.
\end{equation}
The second of these is a primary constraint on our system:
\begin{equation}
    \varphi_0 = p_\Phi \sim 0.
\end{equation}
We use Dirac's notation that a ``$\sim$'' sign indicates that
equality holds weakly; i.e., after all constraints have been
imposed.  The definition of $p_a$ can be rewritten as
\begin{equation}\label{p identity}
\sqrt{F} \cosh \left( \frac{p_a}{a^{d-1}} \right) = \sqrt{
\frac{\dot{a}^2}{\Phi^2} + F }.
\end{equation}
The total Hamiltonian of the model is defined by
\begin{subequations}
\begin{eqnarray}
    H & = & H_g + \alpha_m H_m, \\
    H_g & = & p_a \dot{a} - L_g \\ H_m & =
    & \sum_i \pi_i \dot{\psi}_i - L_m \equiv \Phi a^d {\mathcal H}_m.
\end{eqnarray}
\end{subequations}
Here, we have defined $\pi_i$ as the momentum conjugate to the
matter fields $\psi_i$; i.e., $\pi_i = \di L_m / \di \dot\psi_i$.
Also note our definition of the matter Hamiltonian density
${\mathcal H}_m$. Making use of (\ref{p identity}), we obtain
\begin{equation}\label{H0}
    H = \Phi a^{d-1} \left[ - \sqrt{F} \cosh \left( \frac{p_a}{a^{d-1}}
    \right) + \alpha_m a {\mathcal H}_m \right].
\end{equation}

At this juncture, we should say a few words about the matter
sector of the model. Note that relativistic invariance implies
that all the matter field velocities in the matter Lagrangian must
be divided by the lapse function.  This is because $\Phi\,dt$ is
an invariant but $dt$ by itself is not.  By the same token, $\Phi$
by itself is not an invariant, so we do not expect to see any
appearances of $\Phi$ uncorrelated with a velocity. Therefore,
instead of regarding ${\mathcal L}_m$ as a function of
$\dot{\psi}_i$ and $\Phi$ separately, we can instead regard it as
a function of $v_i = \dot{\psi}_i/\Phi$, which may be thought of
as the proper velocity of matter fields.  We can then define an
alternative Lagrangian density by ${\mathcal
L}_m(\psi_i,\dot\psi_i;a,\Phi) = \bar{{\mathcal
L}}_m(\psi_i,v_i;a)$.  The canonical momenta definition becomes
\begin{equation}\label{canonical momenta}
    \pi_i = a^d \frac{\di\bar{\mathcal L}_m}{\di v_i}.
\end{equation}
We can use this with $\di v_i / \di \Phi = -\dot\psi_i/\Phi^2$ to
deduce that
\begin{equation}\label{A}
    \Lm + \Phi \frac{\di{\mathcal L}_m}{\di \Phi} = -a^{-d}
    \sum_i \pi_i v_i + \Lm = -{\mathcal H}_m.
\end{equation}
Comparing this with (\ref{Lagrange density}) gives the result
\begin{equation}
    \Hm = \rho_\mathrm{tot}.
\end{equation}
This is sensible; the Hamiltonian density of the matter is equal
to its total matter-energy density on solutions.  Since
$\rho_\mathrm{tot}$ is a physical observable it ought to be
independent of the lapse function, which is a gauge-dependent
quantity.  This can be rigorously shown by noting that the
Hamiltonian density may be written as
\begin{equation}\label{B}
    {\mathcal H}_m(\psi_i,\pi_i;a,\Phi) = a^{-d} \sum_i \pi_i v_i -
    \bar{\mathcal L}_m(\psi_i,v_i;a).
\end{equation}
If the definition of the canonical momenta (\ref{canonical
momenta}) is used to replace all instances of $v_i$ with
expressions involving $(\psi_i,\pi_i;a)$ --- which should always
be possible --- this implies that
\begin{equation}\label{C}
    \frac{\di{\mathcal H}_m}{\di
    \Phi} = 0 \quad \Rightarrow \quad \Hm = \Hm(\psi_i,\pi_i;a).
\end{equation}
The last point is that even though we can remove all functional
dependence of $\Hm$ on the proper velocities, we cannot
necessarily find explicit expressions for $v_i =
v_i(\psi_i,\pi_i;a)$.  This will only be possible if
(\ref{canonical momenta}) is invertible, which requires that the
Hessian determinant of the system be non-vanishing:
\begin{equation}\label{hessian}
    \det\left( \frac{\di^2\bar{\mathcal L}_m}{\di v_i \di v_j}
    \right) \ne 0.
\end{equation}
If this fails, the matter Lagrangian is said to be singular and we
will have some number of primary constraints on the matter
coordinates and momenta $\chi^{(1)}_r \sim 0$
\cite{Git90}.\footnote{Late lowercase Latin indices ($r$, $s$,
etc.) run over all matter-related constraints.} One obtains
explicit representations of these by manipulating the system of
equations (\ref{canonical momenta}) to eliminate the velocities,
which means that any primary constraints are independent of the
lapse and $p_a$; i.e., $\chi^{(1)}_r =
\chi^{(1)}_r(\psi_i,\pi_i;a)$.  Can we make the simplifying
assumption that all of the matter Lagrangians that we are
interested in are nonsingular?  The answer is no, largely because
such an assumption would forbid the existence of gauge fields ---
which always involve constraints --- living on the brane, and is
therefore too restrictive.

So, to summarize, we have written down the Hamiltonian of the
model (\ref{H0}) and found out there is at least one primary
constraint $\varphi_0 \sim 0$.  Other primary constraints
$\chi^{(1)}_r \sim 0$ may come from the matter sector.  The next
step is to construct the extended Hamiltonian
\begin{equation}
    H' = H + \mu_0 \varphi_0 + \sum_r \lambda^{(1)}_r \chi^{(1)}_r,
\end{equation}
where $\mu_0$ and $\lambda^{(1)}_r$ are coefficients yet to be
determined. Time derivatives of any quantity are computed through
the usual Poisson bracket:
\begin{equation}
    \dot{A} \sim \{ A, H' \}.
\end{equation}
We now attempt to enforce that the time derivative of $\varphi_0$
is zero.  Making note that both $\Hm$ and $\chi^{(1)}_r$ are
independent of $\Phi$, we see that $\dot\varphi_0 = 0$ implies the
existence of an additional constraint:
\begin{equation}\label{xi def}
    \xi =
    - \sqrt{F} \cosh \left( \frac{p_a}{a^{d-1}} \right) + \alpha_m a
    {\mathcal H}_m \sim 0.
\end{equation}
This gives that our original Hamiltonian is proportional to a
constraint $H = \Phi \xi$ and therefore vanishes weakly, which is
characteristic of reparametrization invariant systems.  The
\emph{first-stage Hamiltonian} of our model is then defined as
\begin{equation}
    H^{(1)} =  \mu_0 \varphi_0 + \mu_1 \xi + \sum_r \lambda^{(1)}_r
    \chi^{(1)}_r,
\end{equation}
where $\mu_1$ is yet another undetermined coefficient.

Now, let us turn our attention to the $\chi^{(1)}$ constraints. We
must demand that each of these is conserved in time, which leads
to the condition $\{\chi^{(1)}_r,H^{(1)} \} \sim 0$.  This in turn
can generate new constraints $\chi^{(2)} \sim 0$, which then
defines a new second stage Hamiltonian $H^{(2)}$.  But then we
need to demand that $\chi^{(2)}$ be conserved, which can generate
even more constraints $\chi^{(3)} \sim 0$.\footnote{For a more
pedagogical account of this procedure, the reader is encouraged to
consult refs.~\cite{Sea03a,Sea03b}.}  Eventually the algorithm
will terminate, say at the $q^\mathrm{th}$ stage, when demanding
that $\{ \chi_r^{(q)},H^{(q)} \} \sim 0$ does not general any new
constraints. Now, by performing a linear transformation on the
$q^\mathrm{th}$ stage matter constraints, we can divide them into
two sets defined by
\begin{subequations}\label{constraint decomp}
\begin{eqnarray}
    \chi^{(q)} & = & \varrho^{(q)} \cup \varpi^{(q)} = (
    \varrho^{(q)}_1, \varrho^{(q)}_2,\dots,\varpi^{(q)}_1,
    \varpi^{(q)}_2,\ldots ), \\
    0 & \sim & \{\varrho^{(q)}_I,\varrho^{(q)}_J \}, \\ 0 & \sim &
    \{\varrho^{(q)}_I,\varpi^{(q)}_R \}, \\ 0 & \nsim &
    \det\{ \varpi^{(q)}_R , \varpi^{(q)}_S \}.
\end{eqnarray}
\end{subequations}
Here, middle and late uppercase Latin indices run over the
$\varrho^{(q)}$ and $\varpi^{(q)}$ constraints respectively.  The
$q^\mathrm{th}$ stage Hamiltonian is then
\begin{equation}\label{M stage H}
    H^{(q)} = \mu_0 \varphi_0 + \mu_1 \xi + \sum_I a^{(q)}_I
    \varrho^{(q)}_I + \sum_R b^{(q)}_R \varpi^{(q)}_R, \quad
    \dot{A} \sim \{ A, H^{(q)} \},
\end{equation}
where the $a^{(q)}$ and $b^{(q)}$ coefficients are undetermined.
Most importantly, it can be shown that all of the members of the
$\varrho^{(q)}$ and $\varpi^{(q)}$ sets are independent of $\Phi$.

We now have the complete set of constraints for our model:
$\varphi_0$, $\xi$, $\varrho^{(q)}$, and $\varpi^{(q)}$. According
to Dirac \cite{Dir64,Git90}, the next step is to categorize them
as first-class and second-class constraints.  It is obvious that
$\varphi_0$ is first-class since we have already established that
it commutes with the other constraints under the Poisson bracket.
It is also obvious that since $0 \nsim \det\{ \varpi^{(q)}_R ,
\varpi^{(q)}_S \}$, the $\varpi^{(q)}$ constraints are
second-class. Furthermore, we have by construction that the
$\varrho^{(q)}$ constraints commute among themselves and the
$\varpi^{(q)}$ set. Also, since the constraint-generating
procedure ends at the $q^\mathrm{th}$ stage, all of the members of
$\varrho^{(q)}$ set must commute with $\xi$, so they are all
first-class constraints. Using these facts allows us to solve for
the $b^{(q)}$ coefficients explicitly by setting
$\dot\varpi_R^{(q)} = 0$:
\begin{equation}\label{b soln}
    b_R^{(q)} = - \sum_{S} \{\varpi^{(q)}_R,
    \varpi^{(q)}_S \}^{-1} \{\varpi^{(q)}_S, \mu_1 \xi \}.
\end{equation}
Here, we have defined $\{\varpi^{(q)}_R, \varpi^{(q)}_S \}^{-1}$
as the matrix inverse of $\{\varpi^{(q)}_S, \varpi^{(q)}_P \}$
such that
\begin{equation}
    \delta_{RP} = \sum_S \{\varpi^{(q)}_R, \varpi^{(q)}_S \}^{-1}
    \{\varpi^{(q)}_S, \varpi^{(q)}_P \}.
\end{equation}
The only thing left is $\xi$ itself. Without knowing more about
the $\varpi^{(q)}$ constraints, we cannot say with certainty that
they do or do not commute with $\xi$ under the Poisson bracket.
(If they did, then $b_R^{(q)} \sim 0$.) But this ignorance is not
really important if we move over to the Dirac bracket formalism.
We define the Dirac bracket between two phase space functions as
\begin{equation}
    \{A,B\}_* = \{A,B\} - \sum_{RS}
    \{A,\varpi^{(q)}_R\} \{\varpi^{(q)}_R,
    \varpi^{(q)}_S \}^{-1} \{\varpi^{(q)}_S, B
    \}.
\end{equation}
So defined, the Dirac bracket has the same basic properties as the
Poisson bracket; i.e., it is antisymmetric in its arguments, it
satisfies the Jacobi identity, etc.  Under the Dirac bracket, each
of the $\varpi^{(q)}$ constraints commute with every phase space
function \emph{strongly}:
\begin{equation}
    \{ A, \varpi^{(q)}_R \} = 0.
\end{equation}
This implies that we can impose $\varpi^{(q)}_R = 0$ as a strong
equality; i.e., we can use the second-class constraints to
simplify the first-class constraints.  Also, the Hamiltonian can
be written in a more streamlined form:
\begin{equation}
    H_\mathrm{tot} = \mu_0 \varphi_0 + \mu_1 \xi + \sum_I
    a^{(q)}_I \varrho^{(q)}_I, \quad \dot{A} \sim \{ A, H_\mathrm{tot}
    \}_*.
\end{equation}
It is easy to confirm that the time evolution equation under the
Dirac bracket using $H_\mathrm{tot}$ is the same as the one under
the Poisson bracket (\ref{M stage H}) if one makes use of (\ref{b
soln}).  Also under the Dirac bracket, both $\xi$ and
$H_\mathrm{tot}$ are realized as first-class quantities. Finally,
the coefficients $\mu_0$, $\mu_1$, and $a^{(q)}$ are completely
arbitrary and hence represent the gauge freedom of the system.

If we wanted to proceed with the Dirac quantization of the model
at this point, we would promote the Dirac brackets to operator
commutators, choose a representation, and then restrict the
physical Hilbert space by demanding that all state vectors be
annihilated by the first-class constraints.  The only impediment
to the immediate implementation of this procedure is the
functional form of $\xi$. As written, $\xi$ contains a hyperbolic
function of $p_a$. This will be problematic if we choose the
standard operator representation $\hat{p}_a = i \di/ \di a$
because the operator $\hat\xi$ will contain $\di / \di a$ to all
orders, essentially resulting in an infinite-order partial
differential equation. There are two ways to remedy this; we could
choose a non-standard operator representation, or we can try to
rewrite the constraint in a different way at the classical level.
Let opt for the latter strategy.\footnote{Koyama \& Soda
\cite{Koy00} have previously considered the quantization of vacuum
branes by modifying the Hamiltonian constraint at the classical
level, but their transformed constraint is somewhat different from
the one we are about to present.} From the theory of constrained
Hamiltonian systems, we know that we can transform one set of
constraints into another set by applying a linear transformation
matrix.  The only requirement is that the matrix be non-singular
on the constraint surface.  In this case, we want to replace a
single constraint $\xi$ with an equivalent constraint $\varphi_1$
such that
\begin{equation}\label{cond 1}
    \xi = 0 \text{ if and only if } \varphi_1 = 0.
\end{equation}
The linear transformation is trivial:
\begin{equation}
    \varphi_1 = \frac{\varphi_1}{\xi} \xi.
\end{equation}
Demanding that the ``transformation matrix'' be non-singular is
equivalent to saying that
\begin{equation}\label{cond 2}
    \frac{\varphi_1}{\xi} \nsim 0;
\end{equation}
i.e., the ratio of the two constraints does not vanish weakly. To
ensure this, it is sufficient to demand that the gradients of
$\xi$ and $\varphi_1$ do not vanish when the constraints are
imposed.  It is straightforward to verify that if we select
\begin{equation}
    \varphi_1 = Fa^d \left[ p_a^2 - a^{2(d-1)} \mathrm{arccosh}^2 \left( \frac{
    \alpha_m a \Hm}{\sqrt{F}} \right) \right],
\end{equation}
then (\ref{cond 1}) and (\ref{cond 2}) are satisfied.  We will
discuss the reason for including the $Fa^d$ prefactor in this new
constraint shortly.  Our final form of the Hamiltonian is then
\begin{equation}
    H_\mathrm{tot} = \mu_0 \varphi_0 + \mu_1 \varphi_1 + \sum_I
    a^{(q)}_I \varrho^{(q)}_I.
\end{equation}
As is appropriate for reparametrization invariant systems, the
Hamiltonian is a linear combination of first-class constraints.

Having completed this short detour, we are ready to quantize the
model.  We make the usual correspondence
\begin{equation}
    [ \hat{A}, \hat{B} ] = i \{ A, B\}_*
    \Big|_{A=\hat{A},B=\hat{B}}.
\end{equation}
One assumption that will make our life easier is
\begin{equation}\label{commutator}
    \{ a, p_a \}_* = 1.
\end{equation}
That is, the Dirac bracket between the conjugate pair $(a,p_a)$ is
the same as the Poisson bracket.  It is possible that this might
not be true, because some of the $\chi^{(2)}$ and later stage
constraints could involve $p_a$.  But for any of the concrete
examples of matter models that we consider, (\ref{commutator})
will hold.  In that case, we make the usual choice of operator and
state representations:
\begin{subequations}
\begin{eqnarray}
    \langle a; \psi_i | \hat{a}|\tilde\Psi\rangle & = & a
    \tilde\Psi(a;\psi_i), \\ \langle a; \psi_i | \hat{\psi}_i |\tilde\Psi\rangle &
    = & \psi_i \tilde\Psi(a;\psi_i) \\ \langle a; \psi_i | \hat{p_a} |
    \tilde\Psi\rangle & = &
    i \frac{\di}{\di a} \tilde\Psi(a;\psi_i).
\end{eqnarray}
\end{subequations}
Here, we consider $|\tilde\Psi\rangle$ to be a possible physical
state vector for our model that is annihilated by the constraints,
while $|a,\psi_i\rangle$ is a state of definite $a$ and $\psi_i$.
Note that since we don't really know much about
$\{\psi_i,\pi_j\}_*$, we have not yet specified and operator
representation of the momenta conjugate to the matter fields
$\hat\pi_i$.  Also, we implicitly assumed that $\tilde\Psi$ is
independent of $\Phi$, which means the constraint $\hat\varphi_0
|\tilde\Psi\rangle = 0$ is satisfied immediately if we select
$\hat{p}_\Phi = i \di/\di\Phi$.  The constraint $\hat\varphi_1
|\tilde\Psi\rangle = 0$ yields the Wheeler-DeWitt equation
\begin{equation}\label{WdW 1}
    \left[ -\frac{\di}{\di a} Fa^d \frac{\di}{\di a} - F a^{3d-2}
    \mathrm{arccosh}^2 \left( \frac{
    \alpha_m a \hatHm}{\sqrt{F}} \right) \right] \tilde\Psi(a;\psi_i) = 0.
\end{equation}
Here, $\hatHm$ is obtained from the classical expression for $\Hm$
by replacing $\pi_i$ by its operator representation; i.e., $\hatHm
= \Hm(\psi_i,\hat\pi_i;a)$. In converting $\varphi_1$ into an
operator, we are faced with two ordering ambiguities.  One of
these is relatively innocuous: Since $\di/\di a$ clearly does not
commute with $Fa^d$, the relative order of the three operators in
the first term on the left is non-trivial.  But by demanding that
the product of these operators be Hermitian, we arrive at the
ordering shown above. The second ordering issue comes from the
fact that we are unsure if
\begin{equation}
    \{a, \Hm\}_* \stackrel{?}{=} 0.
\end{equation}
If this Dirac bracket does not vanish, we have a very serious
problem in the second term of the Wheeler-DeWitt equation.  So we
need to make another assumption, namely that the $\hat a$ and
$\Hm$ operators do indeed commute.  If that is true, then we can
perform a separation of variables by setting
\begin{equation}
    \tilde{\Psi}(a;\psi_i) = \Psi(a)\Upsilon(\psi_i).
\end{equation}
Now, consider the eigenvalue problem associated with the
$\hatHm(\psi_i,\hat\pi_i;a)$ operator where we treat $a$ as a
parameter. Let us select $\Upsilon$ to be an eigenfunction of
$\hatHm$ so that
\begin{equation}
    \hatHm(\psi_i,\hat\pi_i;a) \Upsilon(\psi_i) = \mathcal{U}_m(a)
    \Upsilon(\psi_i).
\end{equation}
Here, the ``eigenvalue'' is $\mathcal{U}_m(a)$ and should, in
general, be labelled by some quantum numbers, as should
$\Upsilon(\psi_i)$. However, in the interests of brevity we will
omit any such decoration.  Since $\Hm$ is classically associated
with the total matter energy density of the matter fields, what we
are essentially doing is finding a basis for the physical state
space in terms of state vectors of definite matter-energy density
$\rho_\mathrm{tot}(a) = \mathcal{U}_m(a)$.

With the total wavefunction partitioned in the way, we can return
to the Wheeler-DeWitt equation.  Since by assumption $\hatHm$ and
$a$ commute, we can expand the second term in (\ref{WdW 1}) in a
series, have $\hatHm$ act on $\Upsilon$ to produce powers of the
eigenvalue, and then collapse the series into a single function.
Then, we can safely divide the resulting reduced Wheeler-DeWitt
equation through by $\Upsilon$. The result is similar to what we
had before:
\begin{equation}
    \left[ -\frac{\di}{\di a} Fa^d \frac{\di}{\di a} - F a^{3d-2}
    \mathrm{arccosh}^2 \left( \frac{
    \alpha_m a \mathcal{U}_m}{\sqrt{F}} \right) \right] \Psi(a) = 0,
\end{equation}
but now we have a purely one-dimensional problem as there are no
references to the matter fields or their conjugate momenta.

Let us return to the rationale for the inclusion of the $Fa^d$
factor.  We could rewrite the Wheeler-DeWitt equation as
\begin{equation}
    \left[ -\frac{1}{a^d\volume} \frac{\di}{\di a} Fa^d\volume
    \frac{\di}{\di a} - F a^{2(d-1)}
    \mathrm{arccosh}^2 \left( \frac{
    \alpha_m a \mathcal{U}_m}{\sqrt{F}} \right) \right] \Psi(a) = 0,
\end{equation}
so that the volume element of the bulk manifold(s) evaluated on
the brane appears in the first term.  This is then equivalent to
\begin{equation}\label{covariant WdW}
    [-\nabla^A \nabla_A + \mathfrak{V} ]\Psi \Big|_{R = a} = 0,
\end{equation}
where $\mathfrak{V}$ and $\Psi$ are scalar functions of the bulk
radial coordinate. In other words, our Wheeler-DeWitt equation is
merely a scalar wave equation in the bulk manifold evaluated at
the position of the brane.  In keeping with the PCP, the
wavefunction does not depend on the spatial coordinates $\theta$,
and in keeping with the static nature of the bulk manifold, the
wavefunction does not depend on the killing time coordinate $T$.
Although we have only established this equality in the special
$(T,\theta^a,R)$ bulk coordinate system, we expect it to hold in
all coordinate systems because (\ref{covariant WdW}) is a
tensorial statement. The inclusion of the $Fa^d$ factor in
$\varphi_1$ is crucial to this conclusion, which is why we have
put it there in the first place.

Let us summarize what has been accomplished in this section.  We
have examined the Hamiltonian dynamics of our model exterior to
the horizon. There are two constraints that come from the
gravitational side of the theory, as is expected by the gauge
invariance of the system. We have also allowed for any number of
constraints that exist among the matter degrees of freedom on the
brane, which means that our model can be used in conjunction with
matter gauge theories. By introducing the Dirac bracket and
transforming one of the constraints from the gravity sector, we
have written the system Hamiltonian as a linear combination of
first class constraints. Employing standard canonical
quantization, we have at the one-dimensional Wheeler-DeWitt
equation:
\begin{equation}\label{final WdW}
    \left[ -\frac{\di}{\di a} Fa^d \frac{\di}{\di a} - F a^{3d-2}
    \mathrm{arccosh}^2 \left( \frac{
    \alpha_m a \mathcal{U}_m}{\sqrt{F}} \right) \right] \Psi(a) = 0.
\end{equation}
The form of the differential operator on the left implies that
this equation is invariant under transformations of the bulk
coordinates.  Here, $\mathcal{U}_m = \mathcal{U}_m(a)$ is an
eigenvalue of $\hatHm(\psi_i,\hat\pi_i;a)$ with respect to the
matter degrees of freedom.  In the process of arriving at
(\ref{final WdW}), we have made the following assumptions:
\begin{enumerate}

\item[(\emph{i})] $\Lm(\psi_i,\dot\psi_i;a,\Phi) = \bar{\mathcal
L}_m(\psi_i,v_i;a)$.

\item[(\emph{ii})] $\{a,p_a\}_* = 1$.

\item[(\emph{iii})] $\{a,\Hm\}_* = 0$.

\end{enumerate}
The first assumption has to do with the relativistic invariance of
of the matter Lagrangian.  The last two have to do with the
structure of the second class constraints associated with the
matter fields; note that these will both be satisfied if the
second class constraints $\varpi^{(k)}$ are independent of $p_a$.

We conclude by commenting on our choice of quantizing our system
with the equivalent constraint $\varphi_1$ instead of the original
constraint $\xi$.  Clearly, it does not matter at the classical
level which of the constraints we use; they both describe the same
dynamics.  However, this is no guarantee that the quantized model
is insensitive to the choice of imposing $\xi \sim 0$ or
$\varphi_1 \sim 0$. In particular, will the physical Hilbert space
defined by $\hat\varphi_1 | \Psi \rangle = 0$ be the same as the
one defined by $\hat\xi | \Psi \rangle = 0$?  Classically, we had
$\xi = \Gamma \varphi_1$ where $\Gamma$ is a phase space function
that does not vanish weakly.  Clearly, if we have the operator
identity $\hat\xi = \hat\Gamma \hat\varphi_1$ the two Hilbert
spaces would be identical.  But there is an ordering ambiguity
here, because we could also have $\hat\xi = \hat\varphi_1
\hat\Gamma$, which would not result in the same Hilbert space
unless $\hat\varphi_1$ and $\hat\Gamma$ commute. This potential
inconsistency --- sometimes called ``quantum symmetry breaking''
--- is endemic in the Dirac quantization programme.  There is
reason to believe that it can be avoided by employing alternative
quantization procedures; for example, if one converts the
first-class constraints in a given system to second-class ones by
adding gauge-fixing conditions, it can be shown that the
associated generating functional for the quantum theory is
invariant under transformations of the constraints
\cite[Sec.~3.4]{Git90}.  However, we should point out that the
difference between $\hat\Gamma \hat\varphi_1$ and $\hat\varphi_1
\hat\Gamma$ is necessarily of order $\hbar$, so we expect the
discrepancy between the physical Hilbert spaces defined by $\xi
\sim 0$ and $\varphi_1 \sim 0$ to be unimportant at the
semi-classical level.

\subsection{The interior region}
\label{sec:positive mass}

When $F(a) < 0$, we have two different actions to choose from for
the Hamiltonization procedure: $S_\pm$.  It turns out that it does
not matter which is employed, they both result in the same
Wheeler-DeWitt equation.  To justify this statement, we will
convert the models described by $S_\pm$ to their Hamiltonian forms
simultaneously.\footnote{It is easy to confirm that if the same
thing were done with $S_\mathrm{tach}$, nothing in the final
result will be different.}  The momentum conjugate to $a$ for the
two actions is
\begin{equation}
    p_a^\pm = \mp a^{d-1}
    \mathrm{arccosh}\left(\frac{\pm\dot{a}}{\Phi\sqrt{-F}}\right).
\end{equation}
Notice that inside the tachyonic region --- where $|\dot{a}| <
\Phi \sqrt{-F}$ --- this momentum becomes imaginary.  This is what
one might expect inside a traditional classically forbidden
region. These expressions for $p_a^\pm$ result in the Hamiltonian
functions:
\begin{equation}
    H_\pm = \Phi a^{d-1} \left[ \pm \sqrt{-F} \sinh \left(
    \frac{p_a}{a^{d-1}} \right) + \alpha_m a \Hm \right].
\end{equation}
Here, $\Hm$ is defined in exactly the same way as before.  For
both Hamiltonians, we still have the primary constraint $\varphi_0
= p_\Phi \sim 0$ representing the time reparametrization
invariance of the model. Demanding that this constraint is
conserved in time yields the secondary constraint(s)
\begin{equation}
    \xi_\pm = \pm \sqrt{-F} \sinh \left(
    \frac{p_a}{a^{d-1}} \right) + \alpha_m a \Hm.
\end{equation}
At this point we need to repeat the constraint generating and
classification procedure described in the previous subsection.
Now, since the first stage matter constraints $\chi^{(1)}$ are
determined entirely by the matter Lagrangian, we expect that they
will be the same inside and outside the horizon.  However, the
higher stage constraints $\chi^{(2,3,\ldots)}$ are obtained by
commuting various expressions with $\xi_\pm$, which means that
there is no reason to believe that those constraints match their
counterparts outside the horizon.  Hence, it is conceivable that
the Dirac brackets derived from the $S$ and $S_\pm$ actions might
be distinct from one another, which may complicate the
quantization procedure.  Such difficulties will be minimized if we
make the assumptions:
\begin{enumerate}

\item[(\emph{i})] $\{a,p_a\}^\pm_* = 1$.

\item[(\emph{ii})] $\{a,\Hm\}^\pm_* = 0$.

\item[(\emph{iii})] $\{\psi_i,\pi_j\}_* = \{\psi_i,\pi_j\}_*^\pm$.

\end{enumerate}
Here, $\{,\}_*^\pm$ are the Dirac brackets defined with respect to
$S_\pm$.  This first two assumptions are the same as the ones made
in the previous subsection carried over to the other side of the
horizon.  The third will ensure that we can choose the same
operator representations for $\hat\psi_i$ and $\hat\pi_i$ inside
and outside the horizon.  At the end of the day, we arrive at the
final Hamiltonian(s)
\begin{equation}
    H^\pm_\mathrm{tot} = \mu_0 \varphi_0 + \mu_1 \xi_\pm + \sum_I
    a^{(q)}_I \varrho^{(q)\pm}_I, \quad \dot{A} \sim \{ A, H_\mathrm{tot}
    \}^\pm_*.
\end{equation}
Here, $\varrho^{(q)\pm}$ is the complete set of matter-related
first class constraints derived from $S_\pm$.  As before, the
Hamiltonian is a linear combination of first class constraints
that we will impose as restrictions on physical state vectors.

The last step before quantizing is to rewrite the $\xi_\pm$
constraints in a more useful form.  It is not hard to see that an
equivalent constraint is
\begin{equation}
    \varphi_1^\pm = Fa^d \left[ p_a^2 - a^{2(d-1)}
    \mathrm{arcsinh}^2\left( \frac{\alpha_m a \Hm}{\sqrt{-F}}
    \right) \right] \sim 0.
\end{equation}
The operator version on this constraints yields the Wheeler-DeWitt
equation inside the horizon.  Notice that there is no sign
ambiguity on the righthand side, which means that both of $S_\pm$
lead to the same wave equation.  To obtain this explicitly, we
make the same choice of representation as we made on the other
side of the horizon.  After separation of variables, we obtain
\begin{equation}\label{final WdW inside}
    \left[ -\frac{\di}{\di a} Fa^d \frac{\di}{\di a} - F a^{3d-2}
    \mathrm{arcsinh}^2 \left( \frac{
    \alpha_m a \mathcal{U}_m}{\sqrt{-F}} \right) \right] \Psi(a) = 0.
\end{equation}
Here as before, $\mathcal{U}_m(a)$ represents an eigenvalue of the
operator $\hatHm(\psi_i,\hat\pi_i;a)$.  Notice that since we chose
the same operator representations as before, $\mathcal{U}_m(a)$
can be taken to be continuous across the horizon.

\subsection{The reduced Wheeler-DeWitt equation and the quantum
potential}\label{sec:limiting potential}

Now that we have obtained Wheeler-DeWitt equations (\ref{final
WdW} and \ref{final WdW inside}) valid inside and outside the
horizon, we can examine how they are stitched together.  The
following definitions are quite useful:
\begin{equation}
    \Psi(a) \equiv \frac{\Theta(a)}{a^{d/2}}, \quad a_* \equiv a_*(a), \quad
    \frac{da_*}{da} \equiv \frac{1}{|F(a)|}.
\end{equation}
The $a_*$ coordinate is the higher-dimensional generalization of
the Regge-Wheeler tortoise coordinate.  Written in terms of these
quantities, the entire Wheeler-DeWitt equation takes the form
\begin{subequations}
\begin{eqnarray}
    0 & = & -\frac{1}{2}\frac{d^2\Theta}{da_*^2} + U(a) \Theta \\ U(a) & = & \frac{1}{2}
    F \left[ \frac{d}{2a} \frac{dF}{da} + \frac{d(d-2)}{4} \frac{F}{a^2} - Fa^{3d-2} W(a)
    \right], \\
    W(a) & = &
    \begin{cases}
        \mathrm{arccosh}^2 \left( \frac{
        \alpha_m a \mathcal{U}_m}{\sqrt{F}} \right), & F(a) > 0,
        \\
        \mathrm{arcsinh}^2 \left( \frac{
        \alpha_m a \mathcal{U}_m}{\sqrt{-F}} \right), & F(a) < 0.
    \end{cases}
\end{eqnarray}
\end{subequations}
On an operational level, this is just a one-dimensional
Schr\"odinger equation with a piecewise continuous potential
$U(a)$ and zero energy.  Note that as usual for covariant wave
equations, the potential is an explicit function of $a$ and
therefore an implicit function of $a_*$.  Also note that if the
bulk is 4-dimensional, the Wheeler-DeWitt equation reduces to the
usual expression for a scalar field around a black hole subjected
to a peculiar potential.

We now discuss some of the properties of the quantum potential
$U(a)$. In this, we restrict our attention to matter fields with
$\mathcal{U}_m > 0$; that is, matter with positive density. We
also assume that $\mathcal{U}_m$ is finite and well behaved for
all $a \in (0,\infty)$.  First and foremost, we are interested in
the behaviour of the potential near a bulk horizon. To gain some
insight, let us assume that all of the zeros of $F$ are simple;
that is, the first derivative of $F$ does not vanish at positions
where $F(a) = 0$, which we denote by $a = a_H$.  Let us focus our
attention on one of these zeros where $F$ is positive to the right
of $a_H$ and negative to the left.\footnote{The reverse of this
case is possible when the bulk has a double horizon; i.e., when $k
= -1$ and $\Lambda |K| < 3/2$. It is straightforward to extend the
argument to this case.} Near $a_H$, we can then approximate
\begin{equation}
    F(a) \approx \mathcal{C} (a - a_H),
\end{equation}
where $\mathcal{C}$ is a positive constant. Now if we take a
closer look at our definition of $a_*$, we see that it actually
defines two separate coordinate patches: one for inside the
horizon $a_*^\mathrm{in}$ and one for outside the horizon
$a_*^\mathrm{out}$.  Then for $a \approx a_H$, we have
\begin{equation}
    F(a) \approx
    \begin{cases}
        +\mathcal{C} \exp(+\mathcal{C} a_*^\mathrm{out}), & a > a_H, \\
        -\mathcal{C} \exp(-\mathcal{C} a_*^\mathrm{in}), & a < a_H.
    \end{cases}
\end{equation}
From this, it is easy to see that the horizon is located at
$a_*^\mathrm{in} = +\infty$ and $a_*^\mathrm{out} = -\infty$. This
then yields
\begin{equation}
    U(a) \approx
    \begin{cases}
        +\frac{\mathcal{C}^2 d}{4a_H} \exp(+\mathcal{C} a_*^\mathrm{out}), &
        a_*^\mathrm{out} \rightarrow -\infty, \\
        -\frac{\mathcal{C}^2 d}{4a_H} \exp(-\mathcal{C} a_*^\mathrm{in}), &
        a_*^\mathrm{in} \rightarrow +\infty.
    \end{cases}
\end{equation}
Clearly, $U(a)$ vanishes at the position of the horizon.
Furthermore, all of the derivatives of $U$ with respect to $a_*$
(``in'' or ``out'') vanish there too.  In other words, the quantum
potential is exponentially flat and completely smooth near any
bulk horizons when expressed as a function of $a_*$.  So as far as
the Wheeler-DeWitt equation is concerned, there are no artifacts
left over from our adoption of a piecewise action; we just have a
one-dimensional Schr\"odinger equation with an analytic potential.

What about the limiting behaviour of $U(a)$ for large and small
$a$? The character of $U(a)$ near the singularity at $a = 0$ is
relatively easy to obtain if we keep in mind that if $W(a)$
diverges as $a \rightarrow 0$, that divergence goes like the
square of a logarithm.  We then obtain:
\begin{equation}
    \lim_{a \rightarrow 0} U(a) = - \frac{1}{2}\left( \frac{Kd}{2a^d}
    \right)^2 \rightarrow -\infty.
\end{equation}
Therefore, the potential is infinitely attractive near the
singularity, even for $K < 0$.  The behaviour for large $a$ is
more complicated, and has to be dealt with on a case-by-case
basis. The calculation is sensitive to both the values of the bulk
parameters and the asymptotic behaviour of the matter density
$\mathcal{U}_m$.  We do not give details here; rather the results
are listed in Table \ref{tab:limits}.  We see that in general,
$U(a)$ diverges to $\pm\infty$ as $a \rightarrow \infty$.  The
fact that $U(a)$ is unbounded from below is not unusual for
quantum cosmological scenarios.  As argued in ref.~\cite{Fei95},
it should be addressed by the specification of boundary conditions
on $\Theta$, which is an issue that we will not consider here.
\begin{table*}
\begin{tabular}{|c|c|c|c|c|}
\cline{4-5} \multicolumn{3}{c|}{}  &
$\lim\limits_{a\rightarrow\infty}
U(a) = +\infty$ & $\lim\limits_{a\rightarrow\infty} U(a) = -\infty$ \\
\hline \multicolumn{3}{|c|}{$\Lambda
> 0$} & $\lim\limits_{a\rightarrow\infty} \alpha_m \mathcal{U}_m <
\sqrt{\frac{2\Lambda}{d(d+1)}}$ &
$\lim\limits_{a\rightarrow\infty} \alpha_m \mathcal{U}_m >
\sqrt{\frac{2\Lambda}{d(d+1)}}$ \\ \hline $\Lambda = 0$ &
\multicolumn{2}{|c|}{$k=+1$} & $\lim\limits_{a\rightarrow\infty}
\alpha_m a \mathcal{U}_m < 1$ &
$\lim\limits_{a\rightarrow\infty} \alpha_m a \mathcal{U}_m > 1$ \\
\cline{2-5} &
\multicolumn{2}{|c|}{$k=-1$} & n/a & for all $\mathcal{U}_m$ \\
\cline{2-5} & $k = 0$ & $K
> 0$ & n/a & for all $\mathcal{U}_m$ \\ \cline{3-5} & & $K < 0$ &
$\lim\limits_{a\rightarrow\infty} \alpha_m a^{\frac{d+1}{2}}
\mathcal{U}_m < \sqrt{-K}$ &
$\lim\limits_{a\rightarrow\infty} \alpha_m a^{\frac{d+1}{2}} \mathcal{U}_m > \sqrt{-K}$ \\
\hline
\end{tabular}
\caption{The large $a$ limits of $U(a)$ for various model
parameters}\label{tab:limits}
\end{table*}

Finally, one can verify using identity (\ref{arccosh identity})
that
\begin{equation}
    W(a)
    \begin{cases}
        > 0, & F(a) < \alpha_m^2 a^2 \mathcal{U}_m^2, \\
        < 0, & F(a) > \alpha_m^2 a^2 \mathcal{U}_m^2.
    \end{cases}
\end{equation}
Notice how the conditions on the right mirror our previous
definitions of classically allowed and classically forbidden
regions (\ref{classically allowed/forbidden}) when $\mathcal{U}_m$
is identified with $\rho_\mathrm{tot}$.  This means that the
contribution to the quantum potential from $W(a)$ is positive in
classically forbidden regions and negative in classically allowed
regions, tending to promote non-oscillatory and oscillatory
behaviour in the brane universe's wavefunction respectively.
However, the terms that do not involve $W(a)$ in $U(a)$ prevent
this from being a hard and fast rule.

To summarize, in this section we have written down the
Wheeler-DeWitt equation for the brane in a form identical to a
Schr\"odinger equation with zero total energy.  We have discussed
the general properties of the potential $U(a)$ appearing in this
equation and shown that Wheeler-DeWitt equation is analytic at the
position of any bulk horizons.  Finally, we saw that the quantum
potential tends to be positive in the non-tachyon classically
forbidden regions identified in Sec.~\ref{sec:classical}, but the
presence of curvature-induced terms muddles this conclusion
somewhat.

\subsection{Perfect fluid matter on the brane}

We now specialize to the case where there is only perfect fluid
matter living on the brane.  For simplicity, we will first assume
that there is only one fluid living on the brane and then make the
trivial generalization to multi-fluid models.

The Lagrangian density for a single irrotational fluid with
equation of state $p = \gamma\rho$ is given in Appendix
\ref{app:dust}.  When this is specialized to our metric
\emph{ansatz} on the brane, we have
\begin{equation}
    {\mathcal L}_m = \frac{1}{2} \left[ e^{(1-\gamma)\vartheta}
    \frac{\dot\psi^2}{\Phi^2} -
    e^{(1+\gamma) \vartheta} \right].
\end{equation}
\emph{A priori}, we see two matter degrees of freedom: $\psi$ and
$\vartheta$.  Note that this Lagrangian meets our relativistic
invariance requirement; i.e., all time derivatives are divided by
$\Phi$.  The conjugate momenta are
\begin{equation}
    \pi_\psi = \frac{\di}{\di \dot\psi} \Phi a^d {\mathcal L}_m = \Phi^{-1}
    a^d e^{(1-\gamma)\vartheta} \dot\psi,
    \quad \pi_\vartheta = \frac{\di}{\di \dot\vartheta} \Phi a^d {\mathcal L}_m  = 0.
\end{equation}
The second of these is the sole primary constraint coming from the
matter sector.  Since it obviously commutes with itself, the
constraint can immediately be classified as one of the $\varrho$
type.  Hence the complete set of first stage constraints from the
matter sector is
\begin{equation}
    \varrho^{(1)} = \pi_\vartheta \sim 0.
\end{equation}
From the expressions for the canonical momenta, $\Hm$ is easily
obtained:
\begin{equation}
    \Hm = \tfrac{1}{2} [ e^{-(1-\gamma)\vartheta} a^{-2d} \pi_\psi^2
    + e^{(1+\gamma)\vartheta} ].
\end{equation}
Now, the next step is to demand that $\varrho^{(1)}$ is conserved
in time. According to the prescription given in the previous two
sections, this involves taking the Poisson bracket of
$\varrho^{(1)}$ with one of $\xi$ or $\xi_\pm$, depending on which
portion of phase space one is working with. Fortunately, we have
that
\begin{equation}
    \{ \varrho^{(1)}, \xi \} = \{ \varrho^{(1)}, \xi_\pm \},
\end{equation}
which means that, for the perfect fluid case, the second stage
constraints associated with each of the brane actions are the
same.  The complete set of matter-related second stage constraints
is
\begin{equation}
    \varpi_1^{(2)} = \pi_\vartheta \sim 0, \quad \varpi_2^{(2)} = \pi_\psi
    - \sqrt\frac{1+\gamma}{1-\gamma} a^d e^\vartheta \sim 0.
\end{equation}
It is easy to verify that these constraints are second class:
\begin{equation}
    \{ \varpi_1^{(2)}, \varpi_2^{(2)} \} = \sqrt\frac{1+\gamma}{1-\gamma} a^d
    e^\vartheta \sim \pi_\psi \nsim 0.
\end{equation}
Since there are no second stage first class constraints there can
be no additional constraints in the system; that is, the
constraint generating procedure terminates after the second stage.
The Dirac bracket structure is easy to write down when there are
only two second class constraints:
\begin{equation}
    \{ A,B \}_* \equiv \{A, B\} + \frac{
    \{A, \varpi_1^{(2)} \} \{ \varpi_2^{(2)}, B\} - \{
    A, \varpi_2^{(2)} \} \{\varpi_1^{(2)}, B\}} { \{
    \varpi_1^{(2)},\varpi_2^{(2)} \} }.
\end{equation}
As mentioned above, within this structure the constraints have
vanishing brackets with everything else in the theory, so we can
realize them as strong equalities $\varpi_1^{(2)} = \varpi_2^{(2)}
= 0$ and thereby simplify $\Hm$ by removing all references to
$\vartheta$:
\begin{equation}\label{fluid H 1}
    \Hm = A_\gamma
    \pi_\psi^{1+\gamma} a^{-d(\gamma+1)}, \quad A_\gamma \equiv
    \frac{ (1-\gamma)^{(\gamma - 1)/2} }{ (1 + \gamma)^{(\gamma
    + 1)/2} }.
\end{equation}
We can tidy this up  by considering the canonical
transformation\footnote{Since the Poisson bracket is invariant
under canonical transformations and the Dirac bracket is defined
with respect to Poisson brackets, the Dirac bracket is also
invariant under canonical transformations.}
\begin{equation}
    Q = \frac{\psi \pi_\psi^{-\gamma}}{A_\gamma (1+\gamma)},
    \quad P = A_\gamma \pi_\psi^{1+\gamma},
\end{equation}
which yields
\begin{equation}\label{H density soln}
    {\mathcal H}_m = P a^{-d(\gamma+1)}.
\end{equation}
This is the matter Hamiltonian to be used for perfect fluids.  It
should be stressed that this form of the fluid Hamiltonian and the
associated Dirac bracket structure is valid throughout the phase
space, both inside and outside the horizon.

A few comments about the perfect fluid Hamiltonian formalism are
in order:  First, the perfect fluid Hamiltonian $\Hm = P
a^{-d(\gamma+1)}$ has been derived directly from Schutz's
variational formalism \cite{Shu70,Shu71} and applied to quantum
cosmology a number of times in the literature
\cite{Lap77,Got83,Lem96,Aca98,Alv98,Bat00,Alv01}.  Our second
comment is that the total Hamiltonian of our model will be
independent of $Q$, which means that $P$ is a classical constant
of the motion.  Therefore, on solutions $\Hm$ evaluates to the
matter energy density of the fluid as a function of $a$ as given
be equation (\ref{rho soln}); i.e., $\Hm = \rho(a)$.  The physical
interpretation of $P$ is the current time fluid density. Our final
comment concerns the following Dirac brackets:
\begin{equation}
    \{a,p_a\}_* = 1, \quad \{a,\Hm\}_* = 0, \quad \{Q,P\}_* = 1.
\end{equation}
The first two equalities mean our perfect fluid matter model
satisfies the assumptions we made in Sec.~\ref{sec:negative mass},
so we can safely use the results derived therein.  The last one
means that we can choose standard operator representations for $Q$
and $P$ when quantizing:
\begin{equation}
    \langle a; Q | \hat{Q}|\tilde\Psi\rangle = Q \tilde\Psi(a;Q),
    \quad \langle a; Q | \hat{P} |\tilde\Psi\rangle = i\frac{\di}{\di Q}
    \tilde\Psi(a;Q).
\end{equation}
Recall that in order to write down the reduced Wheeler-DeWitt
equation, we need to solve the eigenvalue problem associated with
$\hatHm$.  With the operator representation above, that problem is
trivial:
\begin{equation}
    \hatHm \left( Q,i\frac{\di}{\di Q};a \right) \Upsilon(Q) = \mathcal{U}_m(a)
    \Upsilon(Q) \quad \Rightarrow \quad \Upsilon(Q) = e^{-iP_0 Q}, \,\,\,
    \mathcal{U}_m(a) = P_0 a^{-d(\gamma+1)},
\end{equation}
where $P_0$ is a constant.  The rightmost expression can be
directly substituted into equations (\ref{final WdW}) and
(\ref{final WdW inside}) to obtain the wavefunction of the
universe outside and inside the horizon respectively.

These results are easily generalized to multi-fluid models.
Without going into too many details, it should be clear the
Lagrangian density for a multi-fluid model is just the sum of the
Lagrangian densities for each individual fluid.  The
Hamiltonization procedure proceeds in a fashion similar to the
single fluid case, largely because the degrees of freedom for
different fluids do not interact with one another.  Our final
result for the eigenvalue of $\hatHm$ is simply
\begin{equation}
    \mathcal{U}_m(a) = \sum_k P_k a^{-d(\gamma_k + 1)}.
\end{equation}
Here, $k$ is an index that runs over all of the fluids.  The
$k^\mathrm{th}$ fluid has the equation of state $p_k = \gamma_k
\rho_k$ and the constant $P_k$ represents its current epoch
density.  Therefore, for the multi-fluid model the eigenvalue of
$\hatHm$ is the sum of the density associated with each of the
fluids as a function of $a$ and the $P_k$ constants are the
quantum numbers that label the $\Upsilon$ eigenfunction.  Hence,
when we put $\mathcal{U}_m$ into the Wheeler-DeWitt equation and
solve for $\Psi(a)$, we are really solving for state of definite
matter momenta.  In principle, this makes $\Psi(a)$ a member of
basis of matter-momentum eigenstates.

Let us move on to the quantum potential associated with
fluid-filled branes.  Other than the basic limiting behaviour of
$U(a)$ described in the last section, the shape of $U(a)$ is hard
to quantify for completely arbitrary parameter choices. So, in
order to get a feel what the potential really looks like, we plot
it for a wide variety of situations in Fig.~\ref{fig:omnibus}. For
these plots, we take the case of a 5-dimensional bulk $d =3$. The
matter content of the brane is taken to be vacuum energy plus
dust:
\begin{figure*}
\includegraphics{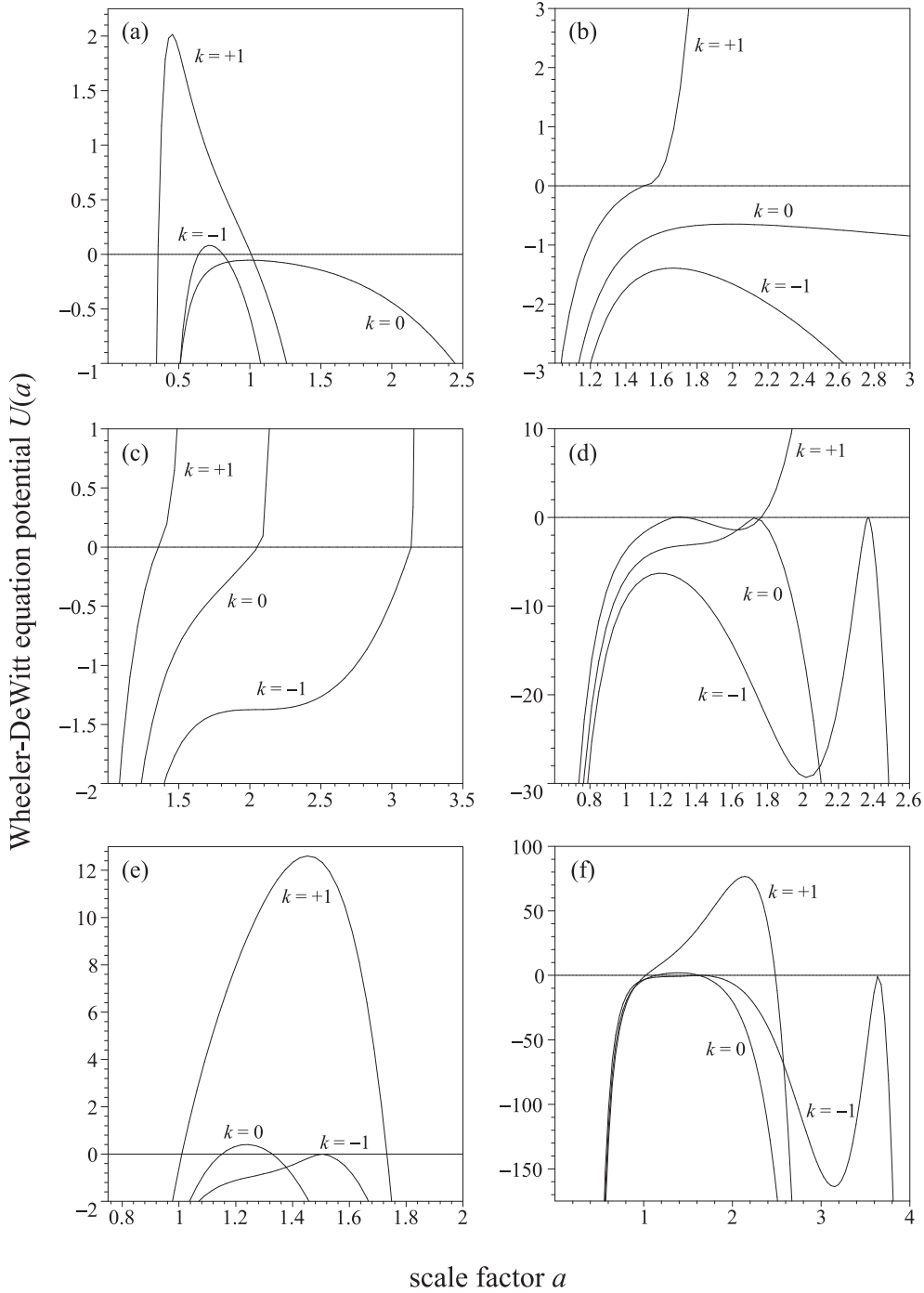}
\caption{The potential in the Wheeler-DeWitt equation for various
model parameters, which are given in Table
\ref{tab:parameter}.\label{fig:omnibus}}
\end{figure*}
\begin{table*}
\begin{tabular}{|c|cccc|}
\hline Panel & $K$ & $\Lambda$ & $\rho_\mathrm{v}$ & $\rho_\mathrm{d}$ \\
\hline \hline (a) & $1/8$ & $0$ & $1/2$ & $1$ \\
(b) & $9/4$ & $0$ & $0$ & $1/2$ \\
(c) & $9/4$ & $3/4$ & $0$ & $1/2$ \\
(d) & $9/4$ & $3/2$ & $1/2$ & $1/2$ \\
(e) & $-9/4$ & $0$ & $3/4$ & $1/16$ \\
(f) & $-9/4$ & $3/8$ & $1/2$ & $1$ \\
\hline
\end{tabular}
\caption{The parameter choices made in each panel of
Fig.~\ref{fig:omnibus}\label{tab:parameter}}
\end{table*}
\begin{equation}
    \alpha_m \mathcal{U}_m = \rho_\mathrm{v} + \rho_\mathrm{d}
    a^{-3}.
\end{equation}
Table \ref{tab:parameter} shows the choice of $K$, $\Lambda$,
$\rho_\mathrm{v}$, and $\rho_\mathrm{d}$ made in each panel of
Fig.~\ref{fig:omnibus}. We now comment on each panel in turn:
\begin{enumerate}

\item[(a)] The $k = +1$ curve in this panel exhibits what seems to
be a finite potential barrier whose left endpoint is the horizon
at $a= \tfrac{1}{\sqrt{8}}$.  Notice also that the $k = -1$
potential also shows a small barrier, which is a purely quantum
effect introduced by the curvature terms in the scalar wave
equation; it has no classical analogue since the classical
potential (equation \ref{energy conservation}) is strictly
negative for this combination of parameters: $V(a) < 0$.

\item[(b)] The $k=+1$ curve in this plot crosses the axis at the
position of the horizon at $a = \tfrac{1}{2}$.  For that case, the
wavefunction can be localized in the interior horizon region.  For
$k =0$ and $-1$, the wavefunction is delocalized over the $a$
axis.

\item[(c)] All potential curves diverge to $+\infty$ in this plot
because the matter density has an asymptotic value less than
$\sqrt{\frac{2\Lambda}{d(d+1)}} = \tfrac{1}{8}$ (see Table
\ref{tab:limits}).  Any apparent ``kinks'' in the potential are
numerical plotting artifacts; the curves are in reality completely
smooth.

\item[(d)] These curves correspond to a marginal case where
$\lim_{a \rightarrow 0} \alpha_m \mathcal{U}_m =
\sqrt{\frac{2\Lambda}{d(d+1)}} = \tfrac{1}{2}$, which is not
included in Table \ref{tab:limits}.  The $k = +1$ curve actually
crosses the zero line three times in this case --- this is
somewhat hard to see without enlarging the plot --- creating a
legitimate potential well.

\item[(e)] The bulk black holes have negative mass in this case.
We see a potential barrier for the $k = 0$ and $k = +1$ case.
There is a horizon at $a = \tfrac{3}{2}$ for $k=-1$, which is
reflected by the vanishing of the hyperbolic potential at that
point.

\item[(f)] All curves in this panel diverge to $-\infty$.  The
$k=-1$ potential goes to zero for two values of $a$, corresponding
to the double horizon structure in this case.  The $k = +1$ and $k
= 0$ potentials show barriers.

\end{enumerate}

To sum up this section: we have specialized the general formalism
presented in previous sections to the case where only perfect
fluids are living on the brane.  The $\hatHm$ operator was seen to
have a very simple form, and its $\mathcal{U}_m(a)$ eigenvalue
merely corresponds to the classical density of the fluids as a
function of $a$.  Also, we have provided plots for a number of
different scenarios corresponding to a 3-brane containing vacuum
energy and dust. Although we have tried to include as wide a
sample of the different types of potentials in
Fig.~\ref{fig:omnibus}, we note that there is actually a
bewildering variety of potential parameter combinations.  So we
must be content with the brief survey above, and we leave a more
systematic study to future work.

\subsection{Tunnelling amplitudes in the WKB approximation}

A number of the panels in Fig.~\ref{fig:omnibus} show that the
curvature singularity at $a = 0$ is hidden behind a potential
barrier.  This suggests the possibility of quantum singularity
avoidance; i.e., the wavefunction of the brane can be engineered
to be concentrated away from the $a = 0$ region.  Or conversely,
we can consider the case where the brane ``nucleates'' by
tunnelling from small to large $a$ in a sort of birth event. At a
semi-classical level, the relevant quantity to both of these
scenarios is the WKB tunnelling amplitude, which gives the ratio
of the height of the wavefunction on either side of the barrier.
If this amplitude is zero (or infinite) we see that the
wavefunction can be entirely contained on one side of the barrier,
while if the amplitude is close to unity it is easy to travel from
one side to the other.

In our situation, the tunnelling amplitude is given by
\begin{eqnarray}\nonumber
    T & = & \frac{ \Psi(a_1) }{ \Psi(a_2) } = \left(\frac{a_2}{a_1}\right)^{d/2}
    \frac{ \Theta(a_1) }{ \Theta(a_2) } \\ \nonumber & \propto & \left(\frac{a_2}{a_1}\right)^{d/2}
    \exp\left( \mp \int_{a_1}^{a_2} da_* \, \sqrt{2 U(a)} \right)
    \\ & = & \left(\frac{a_2}{a_1}\right)^{d/2}
    \exp\left( \mp \int_{a_1}^{a_2} da \, \frac{\sqrt{2 U(a)}}{|F(a)|}
    \right).
\end{eqnarray}
Here, the potential barrier in question is assumed to occupy the
interval $(a_1,a_2)$.  The sign ambiguity in the exponential
allows us to set $\Theta(a_1) \lessgtr \Theta(a_2)$ depending on
what type of scenario we are considering.  Note that we had to
transform the integral of $\sqrt{2U(a)}$ from an integration over
$a_*$ to $a$ at the price of dividing the integrand by $|F(a)|$.
Now, we already know that the potential changes sign whenever $F$
changes signs, so if the bulk contains horizon(s) there will
always be a potential barrier (or barriers) with a horizon as one
of its endpoints.  But using the asymptotic forms of $U$ and $F$
near the horizon developed in Sec.~\ref{sec:limiting potential},
we have that
\begin{equation}
    \frac{\sqrt{2 U(a)}}{F(a)} \propto \frac{1}{\sqrt{F}}, \quad \text{for } F \gtrsim 0.
\end{equation}
Hence, the integrand in $T$ has a pole at one of the integration
endpoints if that endpoint represents a bulk horizon, but the pole
is of order $\frac{1}{2}$.  This leads us to believe that the
integral is actually convergent in such cases, but this is hard to
confirm numerically --- partly because $W(a)$ becomes rather large
near $F(a) \gtrsim 0$.  We would like to report on this phenomena
in the future, but for now let us restrict our discussion to
calculating $T$ for cases without bulk horizons.

\begin{figure}
\begin{center}
\includegraphics{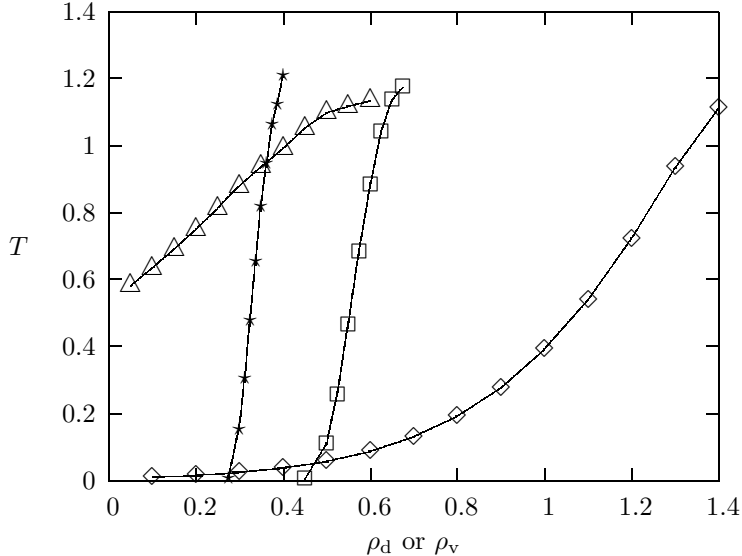}
\caption{The tunnelling amplitude across the potential barrier
when the black hole mass is negative: $K = -\tfrac{9}{4}$.  The
diamonds $(\Diamond)$ indicate the amplitude when $k = +1$,
$\Lambda = 0$, $\rho_\mathrm{v} = \tfrac{1}{2}$ and
$\rho_\mathrm{d}$ is varied; the triangles $(\triangle)$ indicate
the same with $k =0$.  The squares $(\square)$ show $T$ as a
function of $\rho_\mathrm{v}$ when $k = +1$, $\Lambda =
\tfrac{1}{16}$ and $\rho_\mathrm{d} = 1$; the stars $(\star)$ show
the same with $k = 0$.}\label{fig:tunnelling}
\end{center}
\end{figure}

Since the only potential curves in Fig.~\ref{fig:omnibus} that
exhibit barriers not bounded by horizons are associated with
negative mass bulk black holes, we concentrate on the $K < 0$
case.  In keeping with the calculation of the last section, we
again assume that the brane contains vacuum energy and dust, and
$d=3$. In Fig.~\ref{fig:tunnelling}, we plot $T$ versus dust or
vacuum density amplitude for a number of different situations.
Note that the ratio $(a_2/a_1)^{d/2}$ in the definition of $T$
means that we can have $T > 1$.  Two trends are apparent from this
plot: $T$ increases with increasing dust or vacuum density and $T$
is usually smaller for $k = +1$ than for $k = 0$. Recalling that a
small value of $T$ is associated with a high potential barrier,
this means that the brane's wavefunction can be more effectively
localized away from the singularity when its density is low or
when its spatial sections are spherical.  The former is easy to
understand: when the matter on the brane is more dense, it has
more gravitational self energy that promotes collapse.  To
understand why the barriers are higher for $k = +1$, we merely
need to look at the classical potential from equation (\ref{energy
conservation}); the classical potential $V(a)$ increases with
increasing $k$.

\section{Summary and future prospects}\label{sec:summary}

In this paper, we have derived an effective action governing the
dynamics of a matter-bearing boundary wall --- interpreted as a
$(d+1)$-dimensional brane universe --- between a pair of
S-AdS${}_{(d+2)}$ topological black holes in the mini-superspace
approximation.  We found that the configuration space of our model
has a non-trivial structure, and that the action we initially
derived is not real-valued for all classically allowed brane
states.  To get an action that is real for all classically allowed
regions of configuration space, we modified the action in a
piecewise fashion by adding integrals of time derivatives.  There
was one part of configuration space where this procedure failed;
this was the tachyon region where the normally timelike brane is
forced to acquire a spacelike signature.  We then studied the
classical equations of motion for the system in general and
specific cases. We found that the Friedman equation in general
incorporates classically forbidden regions that promote exotic
brane behaviour like big bounces and crunches.  This was confirmed
for a special case that promoted exact analysis. Instanton brane
trajectories were briefly investigated, which led us to conclude
that the tachyon region is not allowed at the classical or
semi-classical level for branes with ordinary matter. The model
was then converted to the Hamiltonian formalism using the
piecewise effective action.  In this procedure, we allowed for
virtually any type of constraint structure among the matter
fields.  Hence, our methods allow for things like gauge fields
living on the brane.  Dirac quantization was accomplished by
rewriting the Hamiltonian constraint on either side of the horizon
in an equivalent form.  Despite the fact that the action is not
analytic everywhere in phase space, the resulting Wheeler-DeWitt
equation is perfectly well behaved for all finite values of the
brane radius.  Furthermore, the differential operator in the wave
equation was shown to be of a form invariant under transformations
of the bulk coordinates.  We finished off by specializing to
perfect fluid matter on the brane and plotting the quantum
potential for a number of different cases.  Where possible, we
calculated WKB tunnelling amplitudes across potential barriers and
discussed their implication for the localization of the brane's
wavefunction away from the cosmological singularity.

We would like to conclude by mentioning a few future projects
based on this work:
\begin{itemize}

\item Our analysis of the quantum potential for 3-branes
containing vacuum energy should be made more systematic. Also,
numerical solutions of the Wheeler-DeWitt equation would be
interesting to pursue, but appropriate boundary conditions on the
universe's wavefunction need to be specified first.

\item There is no essential difficulty involved in generalizing to
electrically charged bulk black holes.  This is of special
interest because a number of recent papers have pointed out that
even a small bulk electric charge can induce exotic classical
brane behaviour, like cyclic universes \cite[for example]{Myu02}.
How such trajectories are realized upon quantization is an open
question.

\item Despite the fact that our formalism allows for general
matter, all of our examples involved perfect fluids. Different
matter fields ought to be considered, especially those that can
give rise to inflation.

\item We can also relax the $\mathbb{Z}_2$ symmetry across the
brane, which allows for the two bulk manifolds to be distinct.
This will be relevant for the ``quantum conchology'' problem
sourced by arbitrary lower-dimensional matter, where one of the
bulk regions is a Minkowski manifold while the other is the
Schwarzschild spacetime.

\end{itemize}
We hope to report on some of these issues in the near future.

\begin{acknowledgments}
We would like to thank Eric Poisson and Paul Wesson for useful
discussions.  SSS would like to thank NSERC and OGS for financial
support.
\end{acknowledgments}

\appendix

\section{Velocity potential formalism for perfect
fluids}\label{app:dust}

In this appendix, we describe our variational principle for
perfect fluid matter matter living on the brane.  Although, the
treatment is somewhat inspired by Schutz's velocity potential
formalism \cite{Shu70,Shu71}, our model is considerably simpler
and is geared towards cosmological applications, not general fluid
configurations.

In general, there may be many fluids living on the brane,
suggesting that the total Lagrangian density is given by a sum
over the Lagrangian densities of the individual fluids:
\begin{equation}
    {\mathcal L}_m = \sum_i {\mathcal L}_{\gamma_i} +
    {\mathcal L}_\mathrm{other},
\end{equation}
where ${\mathcal L}_{\gamma_i}$ is associated with the
$i^\mathrm{th}$ fluid component, which is assumed to have the
equation of state
\begin{equation}
    p_i = \gamma_i \rho_i.
\end{equation}
The contribution ${\mathcal L}_\mathrm{other}$ represents any
non-perfect fluid matter that may be present.

Let us focus in on one of these fluids.  We assume that its
configuration may be described by two scalar potentials  $\psi =
\psi(y^\alpha)$ and $\vartheta = \vartheta(y^\alpha)$.  For the
time being, we have allow the scalars to depend on all of the
$y$-coordinates on the brane, but later we will impose the PCP to
ensure that they depend on time only.  The Lagrangian density and
fluid action are taken as
\begin{subequations}
\begin{eqnarray}
    {\mathcal L}_\gamma & = & - \tfrac{1}{2} [ e^{(1-\gamma)\vartheta}
    h^{\alpha\beta} \di_\alpha \psi \di_\beta \psi +
    e^{(1+\gamma) \vartheta} ], \\
    S_\gamma & = & \frac{\alpha_m}{\V} \int\limits_\Sigma d^{d+1}y
\sqrt{-h} {\mathcal L}_\gamma.
\end{eqnarray}
\end{subequations}
The normalization of $S_\gamma$ is chosen to be consistent with
the effective brane actions $S$, $S_\pm$ and $S_\mathrm{tach}$
derived in Sec.~\ref{sec:action}.  Demanding that the action be
stable with respect to variations of $\vartheta$ yields
\begin{equation}
    h^{\alpha\beta} \di_\alpha \psi \di_\beta \psi = -\left(
    \frac{1+\gamma}{1-\gamma} \right) e^{2\gamma\vartheta}.
\end{equation}
Assuming that $\gamma \in (-1,1)$, we can define a unit timelike
vector directed along the gradient of $\psi$:
\begin{equation}
    u_\alpha = -\sqrt{\frac{1-\gamma}{1+\gamma}} e^{-\gamma\vartheta}
    \di_\alpha \psi, \quad h_{\alpha\beta} u^\alpha u^\beta = -1.
\end{equation}
If we further demand that $\psi$ increases towards the future, we
have that $u^\alpha$ is future-pointing.  Clearly, $u^\alpha$
should be identified as the proper velocity on the fluid.  Note
that because $u^\alpha$ is hypersurface orthogonal, any fluid
described by ${\mathcal L}_\gamma$ must be irrotational.

We can now use equation (\ref{stress energy def}) to obtain the
stress-energy tensor of the matter fields. After simplification,
this reads
\begin{equation}
    T_{\alpha\beta} = \frac{e^{(1+\gamma)\vartheta}}{1-\gamma} [
    (1+\gamma) u_\alpha u_\beta + \gamma h_{\alpha\beta} ].
\end{equation}
Compare this with the stress energy tensor of a perfect fluid
\begin{equation}
    T_{\alpha\beta} = (\rho + p) u_\alpha u_\beta + p h_{\alpha\beta}.
\end{equation}
The two expressions are the same if we make the identifications
\begin{equation}
    \rho = \frac{e^{(1+\gamma)\vartheta}}{1-\gamma}, \quad p =
    \gamma\rho.
\end{equation}
This gives us the density and pressure of our fluid in terms of
the $\vartheta$ field.

We can obtain the final equation of motion by varying the action
with respect to $\psi$, which eventually gives
\begin{equation}
    \nabla_\alpha [ \rho^{1/(1+\gamma)} u^\alpha ] = 0.
\end{equation}
To move further, we need to impose the brane metric ansatz
(\ref{brane metric}) and the PCP, which implies that $u_\alpha = -
\Phi \di_\alpha t$ by isotropy.\footnote{This is equivalent to
demanding that $\psi = \psi(t)$.}  Then, the equation of motion
gives that
\begin{equation}\label{rho soln}
    \rho = \rho_0 a^{-d(1+\gamma)},
\end{equation}
where $\rho_0$ is the fluid density at the current epoch, defined
by $a = 1$.  This is consistent with the first law of
thermodynamics on the brane:
\begin{equation}
    d(\rho a^d) = -p d(a^d).
\end{equation}
Therefore, we have shown that our assumed Lagrangian density
${\mathcal L}_\gamma$ reproduces the stress-energy tensor and
equations of motion of a perfect fluid in a cosmological setting.
A final note, if we evaluate the Lagrangian density on solutions,
we get that ${\mathcal L}_\gamma = p$; i.e., the Lagrangian
density is the pressure of the perfect fluid as in Schutz's work
\cite{Shu70}.

\bibliography{text}

\end{document}